\documentclass[a4paper,11pt]{article}
\usepackage{jheppub} 
\usepackage{lineno}
\usepackage{mathtools}
\usepackage{physics}
\usepackage{amsmath}
\usepackage{float}
\usepackage{tikz}

\newcommand\secref[1]{{\S\ref{#1}}}
\newcommand\appref[1]{{Appendix~\ref{#1}}}
\newcommand\figref[1]{{Figure~\ref{#1}}}
\newcommand\tabref[1]{{Table~\ref{#1}}}

\def\index{{\mathcal{I}}}   
\def\QS{{\mathcal{Q}}}
\def\SQ{{\mathcal{S}}}

\def\NS{{\mathcal{N}}}

\def\Complex{\mathbb{C}}

\def\ClassS{{\text{Class}\;\mathcal{S}}}
\def\trace{{\mathrm{Tr}}}
\def\LieAlg{{\mathfrak{g}}}

\title{Ground States of $\ClassS$ Theory on ADE Singularities and dual Chern-Simons theory}

\author{Emil~Albrychiewicz,}
\author{Andr\'{e}s Franco Valiente,}
\author{Ori~J.~Ganor,}
\author{Chao~Ju}

\affiliation{
Center for Theoretical Physics and Department of Physics,
  University of California,\\
Berkeley, CA 94720, USA., \\
and
Theoretical Physics Group, Lawrence Berkeley National Laboratory, \\
Berkeley, CA 94720, USA}
\emailAdd{ealbrych@berkeley.edu}
\emailAdd{andresfranco@berkeley.edu}
\emailAdd{ganor@berkeley.edu}
\emailAdd{cju19@berkeley.edu}

\abstract{In radial quantization, the ground states of a gauge theory on ADE singularities $\mathbb{R}^4/\Gamma$ are characterized by flat connections that are maps from $\Gamma$ to the gauge group. We study $\ClassS$ theory of type $\mathfrak{a}_1=\mathfrak{su}(2)$ on a Riemann surface of genus $g>1$, without punctures. The fundamental building block of $\ClassS$ theory is the trifundamental Trinion theory - a low energy limit of two M5 branes compactified on the three-punctured Riemann sphere. We show, through the superconformal index, that the supersymmetric Casimir energy of the trifundamental theory imposes a constraint on the set of allowed flat connections, which agrees with the prediction of a duality relating the ground state Hilbert space of $\ClassS$ on ADE singularities to the Hilbert space of a certain dual Chern-Simons theory whose gauge group is given by the McKay correspondence. The conjecture is shown to hold for $\Gamma=\mathbb{Z}_k$, agreeing with the previous results of Benini et al. and Alday et al. A non-abelian generalization of this duality is analyzed by considering the example of the dicyclic group $\Gamma=\text{Dic}_2$, corresponding to Chern-Simons gauge group SO$(8)$.}

\begin{document}
\maketitle
\flushbottom

\section{Introduction}
\label{sec:intro}

There is a duality \cite{Vafa:1994tf, Nakajima:1994nid} between the ground states of $G=\text{U}(q)$, $\NS=4$ supersymmetric Yang Mills theory on $\mathbb{R}^4/\Gamma$ and the ground states of level $q$ Chern-Simons theory \cite{Witten:1988hf} with gauge group $G(\Gamma)$ on $T^2$. Here, $\Gamma$ is a finite subgroup of SU$(2)$, and $G(\Gamma)$ is the gauge group given by the McKay correspondence~\cite{mckay1981graphs}. For SYM in radial quantization, the ground states on  the orbifold $(S^3/\Gamma)\times S^1$ are characterized by nontrivial holonomies of the connection. These holonomies are classified by G-monodromy on the base space or equivalently by group homomorphisms ($\pi_1(S^3/\Gamma)=\Gamma) \rightarrow G$ up to conjugation.  The duality is easy to extend to SU$(q)$ group. Compared with the U$(q)$ case, the theory with SU$(q)$ group has fewer ground states since the image of homomorphisms presenting the flat connections need to be given by a matrix with unit determinant. On the Chern-Simons side, the set of dual ground states correspond to suitable quotients of the weight lattice of $G(\Gamma)$, explained later. 

This paper explores a natural generalization of this idea where, in addition to the ADE singularity, the $N$ M5 branes are compactified on a Riemann surface $\Sigma$ of genus $g>1$. This is the $\ClassS$ construction proposed and explored in~\cite{Gaiotto:2009we,Gaiotto:2009hg} as a compactification of the 6D $\NS=(2,0)$ SCFT \cite{Witten:1995zh} on $\Sigma$. The difference between the
$g=1$ and $g>1$ cases is twofold. First, the $g=1$ case has $\NS=4$ supersymmetry, while the $g>1$ case only has $\NS=2$. Second, the gauge group SU$(2)^{2g-1}$ has more than one SU$(2)$ factor. Naively, if each SU$(2)$ gauge field gives rise to $k$ flat connections on the ADE singularity, one would expect the degeneracy of ground states to be $k^{2g-1}$. However, there is a nontrivial constraint that the flat connections must satisfy arising from minimizing the Casimir energy. We compute this Casimir energy in radial quantization of the Trinion theory (the building block of $\ClassS$ theories) on $S^3/\Gamma$, for the case $N=2$ with either an abelian $\Gamma=\mathbb{Z}_k$ or a dicyclic $\Gamma=\text{Dic}_2$, corresponding to ADE singularities of type $A_{k-1}$ or $D_4$. 
The results for the number of ground states are consistent with the conjecture that the Hilbert space of ground states on $S^3/\Gamma$ is equivalent to the states of Chern-Simons theory on $\Sigma$, with gauge group $G(\Gamma)$ and level $N$. 
The dimension of the Hilbert space of Chern-Simons theory on a Riemann surface is calculated using the fusion coefficients of an associated affine Lie algebra \cite{Witten:1988hf}. The technical observation is that flat connections of SU$(N)$ on $S^3/\Gamma$ correspond to weights of the affine Lie algebra of $G(\Gamma)$ at level $N$ \cite{Vafa:1994tf}, and the Casimir energy condition of the Trinion theory restricts triplets of such weights.
For $\Gamma=\mathbb{Z}_k$, the technical observation is that the Casimir restriction forbids those triplets of weights whose fusion coefficients are zero in the affine Lie algebra, as was previously observed in \cite{Benini:2011nc,Alday:2013rs,Razamat:2013jxa}.

The main new result of this paper is for $\Gamma=\text{Dic}_{2}$, where we verify a similar relation, but with one exception -- there is a triplet of weights whose fusion coefficient vanishes and yet the corresponding Casimir energy is lower. In fact, the state corresponds to an operator of conformal dimension $\Delta<0$. Because it is an operator that cannot move away from the singularity of $\mathbb{R}^4/\Gamma$, it does not necessarily violate unitarity.
We conjecture that a similar connection between Casimir energies and Chern-Simons fusion rules holds for $\ClassS$ theory on other ADE singularities and with other gauge groups for which no simple Lagrangian descriptions exist. However, the existence of the operator with negative conformal dimension creates complications for genus $g>3$. In those cases flat connections can be constructed for which contributions of positive and negative dimensions from all the trinions combine so that the resulting dimension is zero but there is no obvious corresponding state of the dual Chern-Simons theory. For $g>3$ we can only confirm that a subsector of the $\Delta=0$ space of operators matches the space of states of the dual Chern-Simons theory on the same Riemann Surface.

We note that the connection between $\ClassS$ theory on $\mathbb{R}^4/\mathbb{Z}_k$ and Chern-Simons theory at level $k$ has been extended to the famous 3D/3D correspondence \cite{Dimofte:2011py, Gukov:2015sna, Gukov:2016lki} whereby more general partition functions of 6d SCFTs on a product of a lens space $L(k,1)\simeq S^3/\mathbb{Z}_k$ times a 3-manifold $M_3$ (which can be taken in particular to be $\Sigma\times S^1$) can be computed. Thus, the partition function of a 3D theory $T_N[M_3]$ [the low energy limit of the $(2,0)$ theory on $M_3$] is related via the 3d/3d correspondence to (complexified) Chern-Simons theory on $M_3$. But we are not aware of extensions of the 3d/3d correspondence to orbifolds $S^3/\Gamma$ with $\Gamma$ from the D-series.

The outline of the paper is as follows. In section \secref{sec:review} we review the Chern-Simons duality and state the problem. In section \secref{sec:index} we review the computation of the superconformal index within the context of $\ClassS$ theories and the Macdonald limit~\cite{Gadde:2011uv}, and apply those methods to compute the Casimir energies. For the calculation of the superconformal index, we follow~\cite{Benini:2011nc} in the case of $\Gamma=\mathbb{Z}_k$ and provide detailed calculation for the case of $\Gamma=\text{Dic}_2$ in \appref{app:dic2}. In section \secref{sec:counting} we show explicitly the counting of ground states for a non-Lagrangian theory: $N$ M5 branes (SU$(N)$ theory), with $N>2$, placed on a genus two Riemann surface along a $\mathbb{Z}_2$ orbifold singularity. By the level-rank duality on the dual Chern-Simons theory side, the number of ground states for this theory is the same as that of the number of ground states for two M5 branes on the $\mathbb{Z}_N$ singularity. In section \secref{sec:generalization} we conjecture a generalization of this duality to other gauge groups and other ADE singularities and conclude the paper with some leftover open questions.

\section{The duality and the statement of the problem}
\label{sec:review}

Before analyzing the duality for $\ClassS$ theory, we begin by mentioning an observation made in \cite{Vafa:1994tf}, which can be interpreted as a duality between ground states of $U(q)$ $\NS=4$ supersymmetric Yang Mills theory on $\mathbb{R}^4/\Gamma$, where $\Gamma$ is a finite subgroup of SU$(2)$, and level $q$ Chern-Simons theory on $T^2$ with gauge algebra $\mathfrak{g}(\Gamma)$, where $\mathfrak{g}(\Gamma)$ is the Lie algebra from the McKay correspondence \cite{mckay1981graphs}. The duality relies on a result \cite{Nakajima:1994nid} about the moduli space of $U(q)$ instantons on $\mathbb{R}^4/\Gamma$. (See also \cite{Cherkis:2009hpw,Cherkis:2009jm} for the construction of instanton solutions on Taub-NUT spaces.) From now on, we will refer to the duality of \cite{Vafa:1994tf} as the Vafa-Witten (VW) duality. The  subgroups of SU$(2)$ have a Dynkin diagram classification in which the expansion coefficients of the highest root, in terms of simple roots, give the dimensions of the irreducible representations. This McKay correspondence is summarized in \tabref{tab:ADE}.

\begin{table}[ht]
    \centering
    \begin{tabular}{|c|c|c|c|}
       \hline
        $\mathfrak{g}(\Gamma)$ & Dynkin Diagram & $\Gamma$ & $S^3/\Gamma$ \\ \hline \hline
        $\mathfrak{su}_k$ & A$_{k-1}$ & $\mathbb{Z}_k$ & Lens space  \\ \hline
        $\mathfrak{so}_{2(k+2)}$ & D$_{k+2}$ & Dic$_k$ & Prism manifold  \\ \hline
        $\mathfrak{e}_6$ & E$_6$ & 2T & Tetrahedron \\ \hline
        $\mathfrak{e}_7$ & E$_7$ & 2O & Octahedron \\ \hline
        $\mathfrak{e}_8$ & E$_8$ & 2I & Poincar\'e homology sphere \\ \hline 
    \end{tabular}
    \caption{Summary of the McKay correspondence and resulting spherical 3-manifolds. The groups $\Gamma$ are, respectively, cyclic group, dicyclic (binary dihedral) group, binary tetrahedral, binary octahedral, and binary icosahedral \cite{thurston2014three}.}
    \label{tab:ADE}
\end{table}
 
By canonical quantization of Chern-Simons theory on $T^2$, a basis of ground states can be constructed that is in a one-to-one correspondence to points lying on the lattice quotient~\cite{Elitzur:1989nr}:
\begin{equation}
\label{eqn:statelattice}
\frac{\Lambda_w}{W\ltimes q\Lambda_r},
\end{equation}
where $\Lambda_w$, $\Lambda_r$, $W$ are the weight lattice, the root lattice, and the Weyl group, respectively, of $\mathfrak{g}(\Lambda)$.

A simple extension of the VW duality states \cite{Ju:2024} that the ground state Hilbert space of SU$(q)$, $\NS=4$ supersymmetric Yang Mills theory on $S^3/\Gamma$ is the same as that of a truncated Hilbert space of level $q$ Chern-Simons theory on $T^2$ with gauge algebra $\mathfrak{g}(\Gamma)$, where the truncation is such that only states corresponding to points on the root lattice are kept. Therefore, a basis of ground states of the theory corresponds to the set\footnote{In~\cite{Ju:2023umb}, the lattice~\eqref{eqn:statelattice2} is shown to have a close connection to Ehrhart polynomials, and offers a new perspective on the McKay correspondence.}
\begin{equation}
\label{eqn:statelattice2}
\frac{\Lambda_r}{W\ltimes q\Lambda_r}.
\end{equation}
The elements of \eqref{eqn:statelattice2} correspond to certain highest weight representations that can be expressed explicitly in terms of Dynkin labels (the coefficients in an expansion of the highest weight in terms of fundamental weights) as follows.
Since we are considering only simply laced algebras, all roots of the Lie algebra are of the same length (Kac labels are all equal to $1$), and then the corresponding Dynkin labels $(\lambda_1,\lambda_2,\dots,\lambda_r)$, where $r$ is the rank of $\LieAlg(\Gamma)$, are given by solving the condition
\begin{align}
    \label{eqn:Dynkin}
    \sum_{i=1}^r\lambda_i\leq q,
\end{align}
where $q$ is the level.

As an example, consider the SU$(2)$ SYM theory with $\Gamma=\mathbb{Z}_2$. The ground states are classified by flat connections which are represented by group homomorphisms $g: \mathbb{Z}_2\rightarrow \text{SU}(2)$ that satisfy $g^2=\textbf{1}$, since wrapping the Wilson loop around the $\mathbb{Z}_2$ singularity twice makes the loop contractible to a point. There are two solutions:
\begin{equation}
\label{eqn:SU2holExample}
g_1 =
\begin{pmatrix}
1 & 0 \\
0 & 1
\end{pmatrix}, \quad
g_2 =
\begin{pmatrix}
-1 & 0 \\
0 & -1
\end{pmatrix},
\end{equation}
satisfying this condition.

To apply the VW duality, we look at SU$(2)$ Chern-Simons theory at level $2$. As for SU$(2)$, the simple root is twice the fundamental weight. Only two elements lie in the set \eqref{eqn:statelattice}. In terms of the SU$(2)$ Dynkin label, the states correspond to $\lambda_1=0$ and $\lambda_1=2$,
and we propose that they match the holonomies \eqref{eqn:SU2holExample} with $\lambda_1=0$ corresponding to $g_1$ and $\lambda_1=2$ corresponding to $g_2$.

For a less trivial example, consider the ground states of SU$(3)$ SYM theory on a $\mathbb{Z}_3$ singularity. Let $\omega$ denote the third root of unity. The SYM ground states correspond to the holonomies
\begin{equation}
g_1 =
\begin{pmatrix}
1 & 0 & 0 \\
0 & 1 & 0 \\
0 & 0 & 1
\end{pmatrix}, \quad
g_2 =
\begin{pmatrix}
1 & 0 & 0 \\
0 & -1 & 0 \\
0 & 0 & -1
\end{pmatrix}, \quad
g_3 =
\begin{pmatrix}
1 & 0 & 0 \\
0 & \omega & 0 \\
0 & 0 & \omega^2
\end{pmatrix}, \quad
g_4 =
\begin{pmatrix}
\omega & 0 & 0 \\
0 & \omega & 0 \\
0 & 0 & \omega
\end{pmatrix}. \quad
\end{equation}
The SYM ground states are unique up to the Weyl group identification, so we do not have to count additional solutions where non-zero elements are interchanged along the diagonal. 

On the dual Chern-Simons side, we look at the level $3$ states of SU$(3)$ that are also roots. There are four possibilities in terms of Dynkin labels:
\begin{equation}
(0,0), (1,1), (3,0), (0,3),
\end{equation}
where the non-trivial ones are given in terms of simple roots $\alpha_1=(2,-1),\alpha_2=(-1,2)$ by $\alpha_1+\alpha_2, 2\alpha_1+\alpha_2, \alpha_1+2\alpha_2$, which are representatives for equivalence classes of the quotient \eqref{eqn:statelattice2}. 

\begin{table}[htbp]
\centering
\begin{tabular}{|c| cccc|cc|ccccc|}
\hline
\ & 0 &1 & 2 &  3 & 4  &  5 &  6 & 7 &  8 & 9 & \# \\ \hline
D3 & N & N& N& N& D & D & D&D&D& D & $\times$ \\ \hline
M5 & N & N& N& N& N & D & D&D&D&D&N \\ \hline
\end{tabular}
\caption{The string/M-theory setup. N represents Neumann boundary conditions, and D represents Dirichlet boundary conditions. The 10th direction is indicated by \# and does not exist in the type IIB theory. We label this nonexistent boundary condition by an $\times$.}
\label{tab:setup}
\end{table}

The VW duality can also be motivated with a brane construction and AdS/CFT arguments \cite{Maldacena:1997re,Gubser:1998bc,Witten:1998qj}.  Consider a stack of $q$ D$3$-branes on $\mathbb{R}^4/\Gamma$ in the type IIB theory, which we take along directions 0123 (with Euclidean signature). The question of classifying operators of dimension $\Delta=0$ at the origin can be solved, holographically, by identifying and quantizing the degrees of freedom at the fixed points of the orbifold $AdS_5/\Gamma$. This locus of fixed points is one dimensional and extends from the origin of the boundary $\mathbb{R}^4/\Gamma$ into the bulk of $AdS_5$. We refer to the bulk direction as $r$ (identified with direction 5). We would like to argue that these degrees of freedom are equivalent to Chern-Simons theory (at level $q$) on (an auxiliary) $T^2$. This $T^2$ can be realized with a lift to M-theory, that turns the $q$ D$3$-branes into M$5$-branes, as summarized in Table~\ref{tab:setup}. In the IR limit, the lowest derivative term in the 11D supergravity action reduces to a Chern-Simons action on directions $4r\#$, where $\#$ is an extra dimension after the lift to M-theory. The level of Chern-Simons theory is given by the number of M5 branes, which is $q$. (See also \cite{Dijkgraaf:2007sw} for a related construction, and \cite{Ju:2023umb} for more details.)

A natural question to ask is: what occurs if one compactifies the 4\# directions by a punctured, higher genus Riemann surface $\Sigma$ while keeping the ADE singularity along the 0123 directions? What are the ground states of the corresponding dimensionally reduced 4D theory, and how do dualities of $\ClassS$ theory act on them? Such a compactification of $q$ M5 branes, wrapping $\Sigma$, is known as the $\ClassS$ construction~\cite{Gaiotto:2009we, Gaiotto:2009hg}. Therefore, the question is equivalently phrased as finding the ground states of $\ClassS$ theory on ADE singularities. We cannot directly count the ground states at strong coupling, but we can compute the Witten Index, which is independent of coupling, and is very likely to give the actual number of ground states.

A streamlined recap of the $\ClassS$ construction is the following \cite{Akhond:2021xio}: Consider the 6D $\NS=(2,0)$ SCFT characterized by a simply laced Lie algebra $\mathfrak{g}$. We compactify this on a punctured Riemann surface $\Sigma$ with what is known as a partial topological twist of R symmetries and rotational symmetries to compensate for non-zero curvature and preserve $\NS=2$ SUSY in 4D. For each puncture, suitable SUSY preserving boundary conditions on the corresponding fields can be prescribed \cite{Gaiotto:2009we}. In the IR limit, the resulting theory depends only on the complex structure of $\Sigma$. The technique to analyze this theory is to cut $\Sigma$ into the three-punctured spheres connected by tubes. The different ways of cutting are related by dualities, represented by inequivalent pairs of ``pants decompositions'' \cite{Gaiotto:2009we, Gaiotto:2009gz}. The $(2,0)$ theory on the three-punctured sphere in the IR limit is a 4D $\NS=2$ theory with global SU$(q)^3$ flavor symmetry, known as the Trinion theory $T_q$. For $q=2$ the Trinion theory consists of a half-hypermultiplet that transforms as a doublet under each of SU$(2)$. The simplicity of the Trinion theory in the case of $q=2$ is partly due to pseudoreality of the representations. (See \cite{Tachikawa:2015bga} for a review of the Trinion theory.) We will mostly focus on the case $q=2$, with the only exception of \secref{sec:counting}. The $(2,0)$ theory on the long tubes, in the IR limit, consists of a $\NS=2$ vector multiplet with gauge group SU$(q)$. The gauge couplings are determined by the complex structure of $\Sigma$. The remarkable AGT correspondence allows one to compute the partition function of a $\ClassS$ theory on a squashed sphere using 2D Liouville theory and provides a powerful tool to analyze $\ClassS$ theories \cite{Alday:2009aq}. 
Partition functions of 4D $N=2$ gauge theories on ALE spaces $\mathbb{R}^4/\Gamma$ have been computed in \cite{Bonelli:2011kv, Belavin:2011tb} where a relation with correlation functions of 2d super Liouville and Toda-like parafermionic theories was discovered. The partition function of M$5$-branes on $T^2\times\mathbb{R}^4/\mathbb{Z}_k$ was recently computed in \cite{DelZotto:2023rct,DelZotto:2023ryf}.

\begin{figure}[ht]
\centering
\includegraphics[scale=0.2]{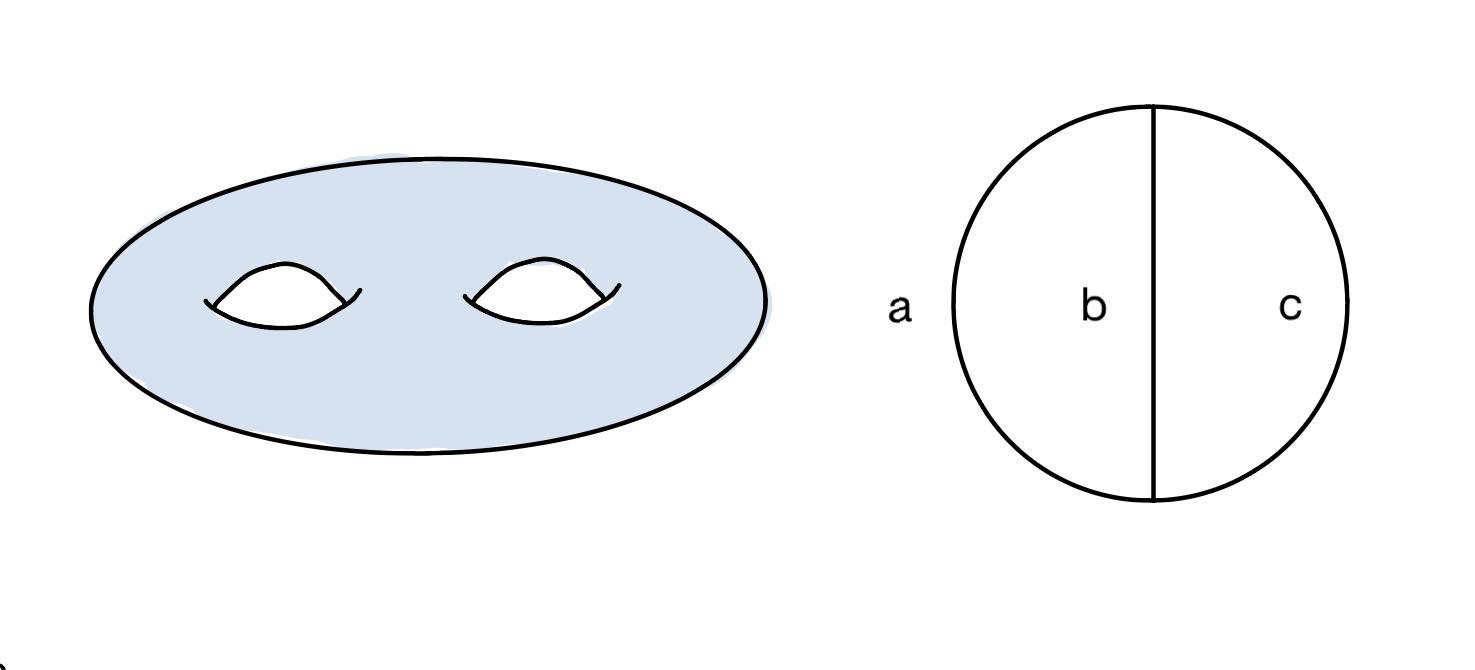}
\caption{Dimensions reduction 6D to 5D to 4D. The intersections of the lines represent the 4D SYM theory. There are three SU$(2)$ gauge fields labeled by a, b, c, respectively. These gauge fields are expected to satisfy certain constraints such that the VW duality holds.
\label{fig:2}}
\end{figure}

In this paper, we quotient out the noncompact directions by an ADE singularity and conformally map, after the IR limit is taken, the smooth portion of $\mathbb{R}^4/\Gamma$ into $\mathbb{R}\times S^3/\Gamma$. We illustrate the compactification using the example of a genus $2$ Riemann surface. This corresponds to two three-punctured spheres glued together and since $q=2$ each puncture is associated with a gauge group SU$(2)$. For demonstration we also take $\Gamma$ to be $\mathbb{Z}_2$. The resulting theory can be analyzed by first compactifying the 6D theory to 5D SYM along one of the basis homology cycles of the genus 2 Riemann surface. The coupling constant of the resulting theory will be proportional to radius of compactification. 
The Riemann surface $\Sigma$ reduces to a one-dimensional space with edges (referred to as {\it propagators}) and vertices where three edges meet (see \figref{fig:2}), and the resulting 5D theory can be reduced to a 4D theory by replacing edges with vector multiplets and vertices with Trinions. 

In \figref{fig:2}, each internal line represents an SU$(2)$ gauge field, labeled by a, b, and c. A basis of ground states of this theory on $S^3/\Gamma$ can therefore be labeled by a triplet
\begin{equation}
(g_a, g_b , g_c),
\end{equation}
where $g_i$ represents the holonomy of gauge field $i$ along the nontrivial cycle of $S^3/\mathbb{Z}_2$. We are looking for group homomorphisms $g_i$ from $\mathbb{Z}_2$ to SU$(2)$, thus each holonomy has to satisfy $g_i^2=\textbf{1}$, i.e. $g_i$ can be either $\text{diag}(1,1)$ or $\text{diag}(-1,-1)$ as in the example \eqref{eqn:SU2holExample}. Therefore, naively, the number of ground states is $2^3=8$. However, the existence of vertices on which the three edges meet suggests that there should be some nontrivial constraint(s) on the gauge field holonomies. 

The constraint can be most easily derived if we assume the VW duality, according to which we expect each holonomy $g_i$ to be dual to a state (also labeled by $i$) of the Chern-Simons theory on $T^2$, or alternatively, a representation of the affine Lie algebra $\mathfrak{g}(\Gamma)$ at level $q$. Based on that, we conjecture that to determine whether a specific ground state $(g_a, g_b, g_c)$ is allowed, one needs to check if the SU$(2)_2$ fusion coefficient $N_{abc}$ is nonzero\footnote{For SU$(2)$, the fusion coefficients can only be either 1 or 0~\cite{kac}.}. The fusion coefficient gives the number of allowed states for Chern Simons theory on $S^2$ with three punctures given by Wilson line insertions (along the time direction) in representations $(a,b,c)$ \cite{Verlinde:1988sn, Witten:1988hf}. Higher genus Riemann surfaces are built by cutting along the basis homology cycles producing three-punctured spheres. We introduce a complete set of states at each cut and then we sew the spheres together based on the fusion rules $N_{abc}$. Although the choice of cuts is not unique, one can show that the result of the gluing is independent of the cuts due to the symmetric properties of the fusion rules. For a genus $2$ Riemann surface, the number of states is given by
\begin{equation}
\label{eqn:NabcNabc} 
{\sum_{a,b,c}}'N_{abc}N^{abc},
\end{equation}
where the sum is over the restricted states, corresponding to elements of \eqref{eqn:statelattice2}. Summing over all elements of the lattice quotient \eqref{eqn:statelattice} reproduces the well known Verlinde formula \cite{Verlinde:1988sn}. We refer to the new constrained sum \eqref{eqn:NabcNabc} as a {\it restricted Verlinde formula.} (For additional details on the Verlinde formula, see for instance \cite{francesco2012conformal}.)

Matching the ground states of $\ClassS$ theory on $S^3/\Gamma$ with states of Chern-Simons theory on $\Sigma$ suggests that the action of dualities of $\ClassS$ theory on ground states correspond to the action of the mapping class group of $\Sigma$ on the states of Chern-Simons theory. More precisely, a duality of $\ClassS$ maps the Hilbert space at one set of coupling constants to another Hilbert space at another set of coupling constants (corresponding to another point in the Teichm\"uller space with the same complex structure). To define an action, we need to connect the two points in coupling constant space (i.e., the Teichm\"uller space) by a path. The path may have a Berry connection, and by ``the action of a duality on ground states'' we mean the combined Berry phase and map between Hilbert spaces. But the exact same structure can be reproduced on the Chern-Simons side, and we conjecture that this is the action that matches. In this paper we only check that the counting of states agrees.

In the SU$(2)$ Dynkin label representation, $a,b,c$ can take the value of either 0 or 2 ($0$ corresponding to the trivial representation and $2$ to the adjoint), and the nonzero $N_{abc}$ are
\begin{align}
\label{eqn:fusionExample}
N_{000}  = 1, \quad N_{022} = 1, \quad N_{220}  = 1, \quad N_{202}  = 1.
\end{align}
We can see that the counting on the Chern-Simons side gives 4 ground states in total, in direct contradiction to the naive answer which is $8$. Therefore, we identify the expectation value of the holonomies of the flat connections, in the appropriate representations, with the dual Chern Simons states as follows:
\begin{align} \nonumber
\begin{pmatrix}
1 & 0 \\
0 & 1
\end{pmatrix} 
\Longleftrightarrow 0, \quad
\begin{pmatrix}
-1 & 0 \\
0 & -1
\end{pmatrix} 
\Longleftrightarrow 2,
\end{align}
the corresponding four ground states on the SYM side can be represented as
\begin{align}  \nonumber
& \left\{ \begin{pmatrix}
1 & 0 \\
0 & 1
\end{pmatrix},
\begin{pmatrix}
1 & 0 \\
0 & 1
\end{pmatrix},
\begin{pmatrix}
1 & 0 \\
0 & 1
\end{pmatrix}
\right\},
\\ \nonumber
& \left\{\begin{pmatrix}
1 & 0 \\
0 & 1
\end{pmatrix},
\begin{pmatrix}
-1 & 0 \\
0 & -1
\end{pmatrix},
\begin{pmatrix}
-1 & 0 \\
0 & -1
\end{pmatrix}
\right\},
\\ \nonumber
&\left\{\begin{pmatrix}
-1 & 0 \\
0 & -1
\end{pmatrix},
\begin{pmatrix}
1 & 0 \\
0 & 1
\end{pmatrix},
\begin{pmatrix}
-1 & 0 \\
0 & -1
\end{pmatrix}
\right\},
\\ \nonumber
&\left\{\begin{pmatrix}
-1 & 0 \\
0 & -1
\end{pmatrix},
\begin{pmatrix}
-1 & 0 \\
0 & -1
\end{pmatrix},
\begin{pmatrix}
1 & 0 \\
0 & 1
\end{pmatrix}
\right\},
\end{align}
following fusion rules \eqref{eqn:fusionExample}.

In the next section, we will explain why the other $4$ combinations of the form $(g_a, g_b, g_c)$ for which $N_{abc}=0$ should not be counted as corresponding to dimension $\Delta=0$ operators -- it is due to a nontrivial Casimir effect that lifts the energy.

We conjecture that this constraint on the holonomies can be generalized to gauge groups SU$(q)$ with $q\ge 3$ by demanding that the supersymmetric index that counts ground states in the corresponding sector is nonzero. Technically, as we will review, these ground states can be counted using the Macdonald limit of the superconformal index that counts $\frac{1}{4}$BPS states (operators), and so the conjecture connects the constraint on fusion rules of three holonomies in Chern-Simons theory to the nonvanishing of the Macdonald limit of a superconformal index. By computing the superconformal index in the Macdonald limit, we will match the Witten index (of dimension $\Delta=0$ operators) with the number of ground states of the corresponding Chern Simons theory for the case of $SU(2)$ and $\Gamma=\mathbb{Z}_k, \text{Dic}_2$.

\section{Casimir energies and the superconformal index}
\label{sec:index}
Counting ground states on the SCFT side can be done through the superconformal index, although the latter obviously contains much more information. 
In our paper, we only need to compute the Casimir energy, so as only to include in the ground state counting those states that correspond to dimension $\Delta=0$ operators, but the technique we apply involves the ``single letter index'', which also is the main step for computing the superconformal index. We therefore start by reviewing the technique for computing the latter.

We compute the $\NS=2$ superconformal index on $S^3/\Gamma\times S^1$, where $\Gamma=\{\mathbb{Z}_k,\, \text{Dic}_2\}$ with gauge group $G=\text{SU}(2)$. For the case of $\mathbb{Z}_k$, the derivation was first performed in \cite{Benini:2011nc} based on the work of 
\cite{Kim:2009wb, Gang:2009qdj, Imamura:2011su}. (Those works explored the generalization to $(S^3/\mathbb{Z}_k)\times S^1$ of the proposed relation \cite{Gadde:2011ik} between partition functions of $\ClassS$ theories on $S^3\times S^1$ and $q$-deformed Yang-Mills theory on $\Sigma$, with the deformation parameter $q$ set to one of the fugacities that define the partition function.)
We will specialize to a particular choice of gauge group rank and ADE subgroup $\Gamma$ at the very end. The important difference between the $\NS=2$ theory on $S^3/\Gamma$ and on $S^3$ is that the former geometry has a nonzero fundamental group: $\pi_1(S^3/\Gamma)=\Gamma$. This allows flat gauge fields to wrap around the singularity, which breaks the gauge group and divides the Hilbert space into sectors labeled by the flat connections. We denote these by homomorphisms $\Gamma\rightarrow G$, up to conjugation.

For $\NS=2$, 4D SCFTs, the superconformal index was first introduced in \cite{Romelsberger:2005eg, Kinney:2005ej}, in close analogy to the Witten index \cite{Witten:1982df} in radial quantization, as:
\begin{align}
    \index=\trace \left\lbrack
    (-1)^Fe^{-\beta\delta}e^{-\mu_iT_i}\right\rbrack,
\end{align}
where the trace is taken over the Hilbert space defined by the spatial hypersurface $S^3$ via the state-operator correspondence, $F$ is a fermion number, and $\delta$ is an anti-commutator of a selected supercharge with its Hermitian conjugate,
\begin{align}   
    \delta=2\{\QS,\QS^\dagger \}.
\end{align}
The index can be shown to be independent of $\beta$, therefore in the limit where $\beta\rightarrow\infty$, only states with $\delta=0$ contribute to the index. 

Fugacities $\mu_i$ were introduced for a complete set of generators $T_i$, which commute with the supercharge $\QS$. The number of these fugacities equals the rank of the subalgebra that commutes with the chosen $\QS$.

The superconformal superalgebra we consider is $\mathfrak{su}(2,2|2)$, and we choose Cartan generators labelled by the simultaneous eigenvalues $(\Delta,j_1,j_2,R,r)$. Here $\Delta$ is the conformal dimension, $(j_1,j_2)$ are the Cartan generators of $\mathfrak{spin}(4)=\mathfrak{su}(2) \otimes \mathfrak{su}(2) $ (related to the double cover rotational symmetry of $S^3$), and $(R,r)$ are generators of the $\mathfrak{su}(2)_R \otimes \mathfrak{u}(1)_r$ R-symmetry group. The supersymmetry charges are denoted by $\QS_{I\alpha}$, $\bar{\QS}^{I}_{\dot{\alpha}}$, $\SQ^{I}_\alpha$, and $\bar{\SQ}_{I\dot{\alpha}}$. The $I$ index is the $\mathfrak{su}(2)_R$ R-symmetry index that ranges from 1 to 2, while the $\alpha=\pm$ and the $\dot{\alpha}=\dot{\pm}$ indices are charges for Cartan elements of $\mathfrak{su}(2)_1$ and $\mathfrak{su}(2)_2$, respectively. In radial quantization, the $\SQ$ and the $\QS$ charges are related by hermitian conjugation:
\begin{equation}
    \SQ_{I\alpha} = (\QS^{I\alpha})^\dagger, \quad \bar{\SQ}_{\;\dot{\alpha}}^I = (\bar{\QS}^{\;\dot{\alpha}}_I)^\dagger.
\end{equation}
Following the conventions of~\cite{Benini:2011nc}, we choose $\QS=\bar{Q}^2_{\;\dot{+}}$ to define the index. The anticommutator relation,
\begin{equation}
\label{eqn:BPS}
    \delta=2\{Q,Q^\dagger\} = \Delta - 2j_2 - 2R + r,
\end{equation}
implies that only states that satisfy the BPS condition
\begin{equation}
\label{eqn:bps}
    \Delta = 2j_2 + 2R - r
\end{equation}
will contribute to the index defined by $\QS$.

The fugacities that are used to refine the index are given by generators of a $\mathfrak{su}(2,1|1)$ subalgebra, which commute with the supercharge $\QS$. The rank of the subalgebra is three, and its charges can be expressed using Cartan elements
\begin{equation}
\label{eqn:charges}
q_1= 2\Delta+2j_2 , \quad q_2=j_1, \quad q_3=r+R,
\end{equation}
in terms of which the index, with fugacities labeled by $t,y,v$, is:
\begin{align}
\label{eqn:TrFull}
    \index=\trace\left\lbrack
    (-1)^Fe^{-\beta\delta}t^{2(\Delta+j_2)}y^{2j_1}v^{-(r+R)}
    \right\rbrack.
\end{align}

We can translate this Hilbert space trace through a formal infinite-dimensional analog of the MacKean-Singer formula and compute the corresponding path integral via supersymmetric localization \cite{Pestun:2016zxk}. The resulting path integral ends up being performed over the conformally flat Euclidean space $S^3\times S^1$. Here $S^1$ is a circle of radius $\beta$. The fermion parity in the operator picture translates to periodic spin structure along the $S^1$ in the path integral, so as to preserve supersymmetry. We skip the explicit details which can be found in the sources mentioned above as well as in many review papers on supersymmetric localization (e.g., \cite{Willett:2016adv}). One arrives at the following formula for the index $\index$:
\begin{align}
\label{eqn:IndexPathInt}
    \index(t,y,v)=\int[dU]e^{-\beta\sum_{\Phi}(-1)^F\frac{E_{\Phi,\pm}}{2}}\exp\left[\sum_{n=1}^{\infty}\frac{1}{n}\hat{\index}\left(t^n,y^n,v^n;e^{i\beta a}\right)\right],
\end{align}
where the ingredients of the integrand are defined as follows.
The integral $\int [dU] (\cdots)$ is performed over the entire gauge group, and $[dU]$ is the Haar measure on the gauge group. But the integrand $(\cdots)$ only depends on the conjugacy class of the gauge group element $U$. The conjugacy class is parameterized by an element $a$ of the Cartan subalgebra, or more precisely, a toroidal subgroup of the gauge group. Any $U$ can be conjugated to such an element, and after conjugation $U$ takes the form $e^{i\beta a}$. The integration variable $a$  can be interpreted as the component of a gauge field along the temporal ($S^1$) direction \cite{Aharony:2003sx}:
\begin{align}
    \label{eqn:GaugeFix}
    \partial_0 a&=\partial_0\left(\frac{1}{V_{S^3}}\int_{S^3} \star A_0\right)=0, \\ 
    \partial_iA^i&=0.
\end{align}

The single letter index that appears in \eqref{eqn:IndexPathInt} is defined as a sum over single particle states, or alternatively, single letter operators as follows. 
Single letter operators are expressed as an arbitrary number of spacetime derivatives acting on a single field. For fields $\Phi$ transforming in the adjoint representation we can expand $\Phi=\sum_{\alpha\neq 0}\Phi^{\alpha}X_\alpha+\sum_{i=1}^{\text{rank}\mathfrak{g}} X_{0,i}\Phi^{0,i}$, where $X_{\alpha\neq 0}, X_{0,i}$ are representatives of a Cartan-Weyl basis of the Lie algebra, and we have $[a,X_\alpha]=\alpha(a)X_\alpha$,
with $\alpha$ denoting nonzero roots of the gauge Lie algebra. Assuming we are also working in a basis of fields that are eigenstates of the quantum numbers $j_1, j_2, \Delta, r, R$, we can write
\begin{align}
    \label{eqn:SingleIndexPhi}
    \hat{\index}(t,y,v;e^{i\beta a})=\sum\left\lbrack
 (-1)^Ft^{2(\Delta+j_2)}y^{2j_1}v^{-(r+R)}
    e^{i\beta\alpha(a)}\right\rbrack,
\end{align}
where the sum replaces the trace of \eqref{eqn:TrFull}, and it is only over single particle states, or alternatively, single letter operators. 

We can also rewrite \eqref{eqn:SingleIndexPhi} as
\begin{align}
    \label{eqn:SingleIndex}
    \hat{\index}(t,y,v;e^{i\beta a})=\sum_{\Phi}(-1)^F e^{-\beta E_{\Phi,\pm}},
\end{align}
where $E_{\Phi,\pm}$ are eigenvalues of the Laplacian and twisted Dirac operators and are given by
\begin{align}
    \label{eqn:EPhiPM}
    E_{\Phi,\pm}=
    -\frac{2}{\beta}(\Delta+j_2)\log t
    -\frac{2}{\beta}j_1\log y
    +\frac{1}{\beta}(r+R)\log v-i\beta\alpha(a).
\end{align}
For the Casimir energy computation, we will need to keep the ratios $(\log y)/\beta$, $(\log t)/\beta$, $(\log v)/\beta$ fixed. (In fact, for our purposes setting $v=y=1$ and $t=\exp\beta$ will be sufficient.)

The primary object of our interest is the expression $\frac{1}{2}\sum_{\Phi}(-1)^F E_{\Phi,\pm}$ that appears in the exponent of the factor multiplying the plethystic exponent \newline
$\exp\sum_{n=1}^{\infty}\frac{1}{n}\hat{\index}\left(t^n,y^n,v^n;e^{i\beta a}\right)$ in \eqref{eqn:IndexPathInt}. We refer to it as the Casimir energy\footnote{The name Casimir energy is a slight misnomer due to the contribution from $j_2$ charge, the more appropriate name would be a twisted or supersymmetric Casimir energy. The discussion regarding the difference can be found in \cite{Kim:2012ava}.} \cite{Kim:2009wb, Imamura:2011su, Benini:2011nc, Kim:2012ava}. By comparing with the single letter index \eqref{eqn:SingleIndex}, we see that
\begin{align}
\label{eqn:CasimirDef}
    \frac{1}{2}\sum_{\Phi}(-1)^F E_{\Phi,\pm}
    =-\frac{1}{2}\lim_{\beta\rightarrow 0}\left[\frac{\partial\hat{\index}}{\partial \beta}\right]_{\text{reg}},
\end{align}
where by $\left[\cdots\right]_{\text{reg}}$, we mean that we regularize the singular contribution, i.e., those terms that diverge in the $\beta\rightarrow 0$ limit [generally behaving as $O(\beta^{-2})$] are removed. To give better intuition of this definition, we provide an example of a simple calculation involving NS fermions in the \appref{app:CasimirExample}. In summary, in order to obtain the Casmir energy and thus some information about the underlying ground states, we need to compute the single letter index.

For $\NS=2$, we have two types of multiplets to consider: the half-hypermultiplet and vector multiplet. The $\NS=2$ vector multiplet consists of a complex scalar $\phi$, two complex Weyl fermions $\lambda^I_{\alpha}, \bar{\lambda}_{I\dot{\alpha}}$ [where we recall that $I=1,2$ is an $SU(2)_R$ R-symmetry index, and $\alpha,\dot{\alpha}$ are 4D spinor indices], and an anti-symmetric two form field-strength $F_{\mu\nu}$, all transforming in the adjoint representation of the gauge group $G$. The $\NS=2$ half-hypermultiplet consists of a complex scalar $q^I$ [a doublet of $SU(2)_R$] and a Weyl fermion $\bar{\psi}_{\dot{\alpha}}$ transforming in fundamental representation of $G$ [which is pseudoreal for our case of $G=SU(2)\times SU(2)\times SU(2)$]. Because this is a ``half-hypermultiplet'', the complex conjugate of the component $q^1$ is identified with $q^2$ (up to charge conjugation of the gauge degrees of freedom), and the complex conjugate of $\bar{\psi}_{\dot{\alpha}}$ constitutes the opposite chirality spinor. All fields become free fields in the localization limit, where the gauge coupling constant is set to zero.

Only those component fields for which $\delta=\Delta-2j_2-2R+r$ vanishes contribute to the single letter index. 
In \tabref{tab:SingleLetter} we list all relevant fields together with their charges and their contributions to this index. The index receives contributions from modes of these fields on $S^3/\Gamma$, and with the help of the state-operator correspondence, the higher modes correspond to single letter operators with insertions of partial derivatives. The operators with derivative insertions are descendants that can be expressed as commutators with the supercharges thanks to the anticommutation relations $\{Q^I_\alpha,\bar{Q}_{J\dot{\alpha}}\}=P_{\alpha\dot{\alpha}}\delta^I_J$. But there are only two spatial derivatives that have $\delta=0$, and only those derivatives create excitations that give a nonzero contribution to the index. The others are canceled by superpartners. 

\begin{table}[ht]
    \centering
    \begin{tabular}{|c|c|c|c|c|c|c|} \hline
        Operators & $\Delta$ & $j_1$ & $j_2$ & $R$ & $r$ & $\hat{\index}$  \\ \hline \hline
        $\phi$ & $1$ & $0$ & $0$ & $0$ & $-1$ & $t^2v$ \\ \hline  
        $\lambda^1_{\pm}$ & $\frac{3}{2}$ & $\pm\frac{1}{2}$ & $0$ & $\frac{1}{2}$ & $-\frac{1}{2}$ & $-t^3y, -t^3y^{-1}$ \\ \hline 
        $\bar{\lambda}_{1\dot{+}}$ & $\frac{3}{2}$ & $0$ & $\frac{1}{2}$ & $\frac{1}{2}$ & $\frac{1}{2}$ & $-t^4$ \\ \hline
        $\bar{F}_{\dot{+}\dot{+}}$ & $2$ & $0$ & $1$ & $0$ & $0$ & $t^6$ \\ \hline 
        $\partial_{\alpha\dot{+}}\lambda^{1\alpha}=0$ & $\frac{5}{2}$ & $0$ & $\frac{1}{2}$ & $\frac{1}{2}$ & $-\frac{1}{2}$ & $t^6$ \\ \hline \hline
        $q^1$ & $1$ & $0$ & $0$ & $\frac{1}{2}$ & $0$ & $t^2v^{-1/2}$ \\ \hline
        $\bar{\psi}_{\dot{+}}$ & $\frac{3}{2}$ & $0$ & $\frac{1}{2}$ & $0$ & $-\frac{1}{2}$ & $-t^4v^{1/2}$ \\ \hline \hline
        $\partial_{\pm\dot{+}}$ & $1$ & $\pm\frac{1}{2}$ & $\frac{1}{2}$ & $0$ & $0$ & $t^3y, t^3y^{-1}$ \\ \hline
    \end{tabular}
    \caption{Summary of operators contributing to the Index (satisfying $\delta=\Delta-2j_2-2R+r=0$) with their charges under symmetry generators. The first part of the table refers to $\NS=2$ vector multiplet components plus Dirac equation for $\lambda^1_\alpha$, the middle part gives $\NS=2$ half-hypermultiplet components and the last one gives contribution from spacetime derivatives.}
    \label{tab:SingleLetter}
\end{table}

If we were only interested in computing the Index on $S^3\times S^1$, the index instructs us to trace over all states satisfying the BPS condition. However, in this paper, we consider the index on an orbifold $S^3/\Gamma\times S^1$, thus some of the states within the Hilbert space are projected out by the orbifold action. This action has no fixed points.
Since $\Gamma$ is a finite subgroup of SU$(2)$, we can embed it within one of the two SU$(2)$ subgroups of the isometry group [$\text{SU}(2)_1\times\text{SU}(2)_2$] of $S^3$.
Since the supercharge \eqref{eqn:BPS} was chosen to commute with SU$(2)_1$ we conveniently embed $\Gamma\subset \text{SU}(2)_1$. 
Modding $S^3$ out by $\Gamma$ also allows for nontrivial flat connections due to the nontrivial fundamental group $\pi_1(S^3/\Gamma)=\Gamma$.
The $\text{SU}(2)_R\times \text{U}(1)_r$ symmetry bundle is trivial over $(S^3/\Gamma)\times S^1$.

Consider a field $\Phi_{j_1,j_2}$ with SU$(2)_1$ and SU$(2)_2$ quantum numbers $j_1, j_2$. It can be extended to a field $\Phi_{j_1,j_2}(x)$ on $S^3\times S^1$ that satisfies, for all $g\in\Gamma$, the invariance condition ${}^g\Phi_{j_1,j_2}({}^gx)=\Phi_{j_1,j_2}(x)$, where ${}^gx$ denotes the geometrical action of $g$ on $x$ following from the orbifold action of $\Gamma$ on the base space $S^3$, and ${}^g\Phi$ denotes the combined action of $g$ on spin and gauge, the latter being nontrivial for a nontrivial flat connection. Using the state-operator correspondence, we map these modes to operators at the origin of $\mathbb{R}^4$ of the form $(\partial\cdots\partial)\Phi_{j_1,j_2}(0)$, with a single field $\Phi_{j_1,j_2}$ and a finite number (possibly zero) of derivative insertions that are invariant under the combined gauge and spin action of all $g\in\Gamma$. The geometric action of $g$ depends only on $j_1$ since $\Gamma$ is embedded only in the SU$(2)_1$ factor. From the elements that contribute to the single letter index (see \tabref{tab:SingleLetter}), the derivatives $\partial_{\pm\dot{+}}$ and the fermionic fields $\lambda^1_{\pm}$ of the vector multiplet are charged under $j_1$, and will contribute to a nontrivial (geometric) transformation of the single letter operator $(\partial\cdots\partial)\Phi_{j_1,j_2}(0)$ under $g$. Finally, we have to keep track of the fact that expressions of the form $(\partial\cdots\partial)\lambda$ that vanish by virtue of the Dirac equation $\partial_{\alpha\dot{+}}\lambda^{1\alpha}=0$ should not be counted in the index.

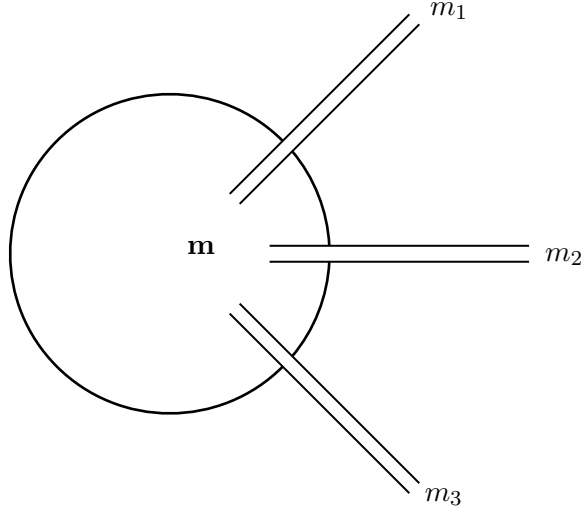
\begin{figure}[ht]
        \centering
        \tikzset{every picture/.style={line width=0.75pt}}
    \begin{tikzpicture}[x=0.75pt,y=0.75pt,yscale=-1,xscale=1]

\draw[line width=1pt] (3:80) arc(3:40:80);
\draw[line width=1pt] (45:80) arc(45:315:80) node[right, xshift=-40, yshift=-40]{$\mathbf{m}$}; 
\draw[line width=1pt] (320:80) arc(320:357:80);

\draw (30,30) -- (120,120);
\draw (35,25) -- (125,115) node[right, xshift=-2, yshift=-5] {$m_3$};

\draw (30,-30) -- (120,-120);
\draw (35,-25) -- (125,-115) node[right, xshift=0, yshift=5] {$m_1$};

\draw (50,-4) -- (180,-4);
\draw (50,4) -- (180,4) node[right, xshift=2, yshift=2] {$m_2$};

    \end{tikzpicture}
        \caption{Three punctured sphere with cut propagators. Holonomies are marked according to the prescription explained in the text.}
        \label{fig:trinionSketch}
    \end{figure}

The path integral is over the moduli space of flat connections. Since $\pi_1(S^3/\Gamma)=\Gamma$ is a finite group, the path integral decomposes into a finite sum over topological sectors with fixed flat connections. We recall that to build a higher genus Riemann surface $\Sigma$ we use a Trinion with three tubes attached as a fundamental block. For the Trinion with three tubes (propagators), we will denote those flat connection sectors by a 3-tuple: $\mathbf{m}=(m_1,m_2,m_3)$, as shown on \figref{fig:trinionSketch}, where $m_i$ labels the flat connection on the $i^{th}$ tube (with $i=1,2,3$ referring to the tube number), and each $m_i$ runs from $0$ to the number of nontrivial flat connections on $S^3/\Gamma$, with $m_i=0$ corresponding to the trivial flat connection. The index becomes 
\begin{align}
\label{eqn:FullIndex}
    \index=\sum_{\mathbf{m}}\int[dU]e^{-\beta\sum_{\Phi}(-1)^F\frac{E_{\Phi,\pm}}{2}}\exp\left[\sum_{n=1}^{\infty}\frac{1}{n}\hat{\index}\left(t^n,y^n,v^n;e^{in\beta a}\right)\right].
\end{align}
We will now provide more details about the flat connections $\mathbf{m}$, and their corresponding Casimir energies. We only need $\hat{\index}$ for the computation of the Casimir energy. For clarity, we will denote the part of \eqref{eqn:FullIndex} that contains the Casimir energy contribution by
\begin{align}
    I_0(\mathbf{m})=e^{-\beta\sum_{\Phi}(-1)^F\frac{E_{\Phi,\pm}}{2}}.
\end{align}

\subsection{Counting holonomy sectors for the Lens space}

\label{sec:lens}

We start by showing that the number of topological sectors labeled by holonomies matches the number of ground states of the corresponding Chern Simons theory on $\Sigma$. A general Lens space is a result of an orbifold action of $\Gamma=\mathbb{Z}_k$ on $S^3$ and is denoted by $L(k,p)$ where $k,p$ are coprime integers. We focus on the case where $p=1$, for which the action of $\mathbb{Z}_k$ can be restricted to act on the $S^1$ fiber of the Hopf fibration over $S^2$. To describe the orbifold action, we parametrize $S^3$ with complex coordinates $(z_1,z_2)\in \Complex^2$ that satisfy $|z_1|^2+|z_2|^2=1$, with the quotient identifying points as
\begin{align}
    \label{eqn:LensSpaceIden}
    (z_1,z_2)\sim(e^{2\pi i/k}z_1, e^{-2\pi i/k}z_2).
\end{align}

Since our primary interest is describing holonomies on the Lens space, we switch gears to discuss how this orbifold action affects the parallel transport (by the flat connection) of fields $\Phi$ transforming in a representation $\rho$ of the gauge group. 
A flat connection $\mathbf{m}=(m_1, m_2, m_3)$ can be represented by homomorphism $\mathbb{Z}_k\rightarrow G=\text{SU}(2)^3$ from $\pi_1(S^3/{\mathbb{Z}_k})\simeq\mathbb{Z}_{k}$ to the gauge group $G$, with $m_i$ corresponding to the $i^{th}$ factor $\Gamma\rightarrow \text{SU}(2)$.
For a given nontrivial flat connection, we choose a gauge such that its holonomy $g$ (along a Hopf fiber) lies in the same toroidal subgroup as $e^{i\beta a}$ [defined above \eqref{eqn:GaugeFix}]. Since $g$ satisfies $g_m^k=1$, up to conjugation we can 
set
\begin{align}
\label{eqn:GDefSU2}
    g_m=\begin{pmatrix}
        e^{2\pi i m/k} & 0\\
        0 & e^{-2\pi i m/k}
    \end{pmatrix},
\end{align}
with $m$ in the range
\begin{align}
\label{eqn:mRange}
    m=0,1,2,\dots,\left[\frac{k}{2}\right],
\end{align}
where by the bracket $\left[k/2\right]$ we mean the largest integer smaller than $k/2$. This identifies $g_m$ in \eqref{eqn:GDefSU2} uniquely up to Weyl conjugation $m\rightarrow -m$. Thus $\mathbf{m}$ can be identified with an element of the Cartan subalgebra of $\mathfrak{g}$. [The identification is ambiguous because of the Weyl conjugation, but we fixed the ambiguity arbitrarily in  \eqref{eqn:mRange}.]

Depending on whether the field $\Phi$ belongs to a half-hypermultiplet or a vector multiplet, it will transform in different representations $\rho$ under parallel transport. For the half-hypermultiplet, $\Phi$ will transform in the fundamental representation by left group multiplication $g_m\Phi$. For the vector multiplet, $\Phi$ will be in the adjoint representation and it will transform under the adjoint action $g_m\Phi g_m^{-1}$. Writing the components of $\Phi$ in a Cartan-Weyl basis, as $\Phi_\rho$ (with $\rho$ a weight of $\mathfrak{g}$), and denoting $\rho(\mathbf{m})$ the action of a weight $\rho$ on the Cartan element $\mathbf{m}$, the effect on the gauge fibers due to parallel transport can be summarized as
\begin{align}
    \Phi \rightarrow e^{2\pi i \rho(\mathbf{m})/k}\Phi.
\end{align}

So far we only described the effect of parallel transport on the gauge degrees of freedom; we now analyze the geometric effect of parallel transport on the field components. Since only the $S^1$ fiber is affected by the orbifold action, we can relate the local coordinates on $S^3$ to the local trivialization of the Hopf fibration via
\begin{align}
    z_1&=e^{i\frac{\xi_1+\xi_2}{2}}\sin{\eta} \\ \nonumber
    z_2&=e^{i\frac{-\xi_1+\xi_2}{2}}\cos{\eta},
\end{align}
where $0\leq\eta\leq\pi/2$, $0\leq\xi_1\leq 2\pi$ and $0\leq\xi_2\leq 4\pi$. The orbifold identification \eqref{eqn:LensSpaceIden} imposes $\xi_2\rightarrow \xi_2+\frac{4\pi}{k}$.

To compare the gauge and geometric factors, we can Fourier decompose $\Phi$ using quantum number $l$ of the $j_1$ generator along the $S^1$ fiber
\begin{align}
    \Phi=\sum_{l}\Phi_l e^{i\xi_2 l}.
\end{align}  
The derivatives $\partial_{\pm\dot{+}}$ have a nontrivial $j_1$ charges, which are $\pm\frac{1}{2}$, respectively.
With additional derivative insertions, $\Phi\rightarrow\partial_{+\dot{+}}^{n_1}\partial_{-\dot{+}}^{n_2}\Phi$, the quantum number $l$ is increased by $\frac{n_1-n_2}{2}$. 

The relevant fields (for the calculation of the index) should be single-valued under parallel transport. With this being understood implicitly, we will not include the derivatives explicitly. 
We impose that the field should be single-valued under parallel transport and hence, we obtain the constraint
\begin{align}
    \Phi\left(\xi_2+\frac{4\pi}{k}\right)=e^{2\pi i \rho(\mathbf{m})/k}\Phi(\xi_2),
\end{align}
which gives a restriction on the numbers of derivative insertions,
\begin{align}
\label{eqn:ProjCond}
    n_1-n_2\equiv\rho(\mathbf{m}) {\pmod k}.
\end{align}
This equation imposes a constraint on the allowed $n_1$ and $n_2$, given the holonomy $\mathbf{m}$ and the weight $\rho$ of the representation of the field component. Thanks to \eqref{eqn:ProjCond}, the Casimir energy for the Trinion, which receives contributions for the hypermultiplets and vector multiplets, depends on $\mathbf{m}$ ~\cite{Benini:2011nc}. The values of $\mathbf{m}$ for which this Casimir energy is minimal, corresponding to the ground states, were derived in \cite{Alday:2013rs}, and we will review the solution below in \eqref{eqn:constraint}. 

At this point, let us recall the definition of the Macdonald limit \cite{Gadde:2011uv}. It is convenient to take a linear combination of the three commuting charges~\eqref{eqn:charges} and define fugacities $p,q,t$ using these new charges. In terms of their eigenvalues, these new charges $q_1', q_2', q_3'$ read:
\begin{align*}
    q_1' &= \frac{q_1+3q_2-2q_3}{3} = j_1+j_2-r, \\
    q_2' &= \frac{q_1-3q_2-2q_3}{3} = -j_1+j_2-r, \\
    q_3' &= q_3 = r+R.
\end{align*}
We redefine the superconformal index using the primed charges
\begin{align}
\label{eqn:IndexNewFug}
    \index &= \tr(-1)^F p^{q_1'}q^{q_2'} t^{q_3'} \nonumber \\
    &= \tr(-1)^F p^{j_1+j_2-r }q^{-j_1+j_2-r} t^{r+R}.
\end{align}
The Macdonald limit is $p\rightarrow 0$.

For the time being we will assume $p>0$ and start by computing the Casimir energy from the single letter index, according to \eqref{eqn:CasimirDef}. We will discuss the $p\rightarrow 0$ limit afterwards.
We need the contribution from the trifundamental half-hypermultiplet $I_0^{H}(\mathbf{m})$, and the contribution $I_0^{V}(m_i)$ from the $i$th propagator ($i\in\{1,2,3\}$). Using the formula \eqref{eqn:CasimirDef}, the single letter contributions from each field to the index \eqref{eqn:IndexNewFug}, and the projection condition \eqref{eqn:ProjCond}, the contributions to the Casimir energy were computed in \cite{Benini:2011nc,Alday:2013rs, Razamat:2013jxa, Razamat:2013opa}, and the results are
\begin{equation}
\label{eqn:casimir1}
    I_0^H(\textbf{m}) = \left(\frac{\sqrt{t}}{pq}\right)^{-\frac{1}{4}\sum_{\mathbf{s}}([\textbf{m}\cdot\textbf{s}]_k - [\textbf{m}\cdot\textbf{s}]^2_k/k)}, \quad I_0^V(m_i) = \left(\frac{\sqrt{t}}{pq}\right)^{[2m_i]_k-[2m_i]^2_k/k},
\end{equation}
where $[x]_k$ means $x{\pmod k}$, $\mathbf{s}=(\pm 1,\pm 1,\pm 1)$ is a vector of possible weights of the trifundamental fields, and the formula for $I_0^H$ contains a sum over all $8$ possibilities for $\mathbf{s}$. 
The powers $-\frac{1}{4}\sum_{\mathbf{s}}([\textbf{m}\cdot\textbf{s}]_k - [\textbf{m}\cdot\textbf{s}]^2_k/k)$ and $[2m_i]_k-[2m_i]^2_k/k$ in \eqref{eqn:casimir1} are proportional to the contributions of the corresponding multiplet to the Casimir energy, but the expressions include two subtractions: (i) an infinite $O(\beta^{-2})$ contribution was removed as discussed after \eqref{eqn:CasimirDef}, and (ii) the ``bare'' Casimir energies were reduced by a finite $\mathbf{m}$-independent amount. Subtraction (ii) is motivated by an expected shift in dimensions of operators on $\mathbb{R}^4/\Gamma$ relative to Casimir energies on $S^3/\Gamma$ (computed via the state-operator correspondence), similar to the $-c/24$ shift in the context of 2D CFTs. The powers in \eqref{eqn:casimir1} correspond to dimensions of operators, rather than energies per se.

Based on the results above, we can write the final answer for the Trinion by multiplying the contributions from the hypermultiplet and vector multiplets:  
\begin{equation}
\label{eqn:casimir2}
    I_0(\mathbf{m}) = I_0^H(\textbf{m}) \sqrt{I_0^V(m_1)I_0^V(m_2)I_0^V(m_3)}, 
\end{equation}
where the square root is because each tube (propagator) is shared by two trifundamental vertices, and \eqref{eqn:casimir2} was defined as the contribution of only a single Trinion. 

Explicitly:
\begin{align}
\label{eqn:I0m}
    I_0(\mathbf{m})=\left(\frac{pq}{\sqrt{t}}\right)^{\frac{1}{4}\sum_{\mathbf{s}}([\textbf{m}\cdot\textbf{s}]_k - [\textbf{m}\cdot\textbf{s}]^2_k/k)-\frac{1}{2}\sum_{i}\left([2m_i]_k-[2m_i]^2_k/k\right)}.
\end{align}
The expression
\begin{equation}
\label{eqn:CasimirResultZk}
\frac{1}{4}\sum_{\mathbf{s}}([\textbf{m}\cdot\textbf{s}]_k - [\textbf{m}\cdot\textbf{s}]^2_k/k)-\frac{1}{2}\sum_{i}\left([2m_i]_k-[2m_i]^2_k/k\right)
\end{equation}
is the sought after Casimir energy. It turns out to be non-negative for any tuple $\mathbf{m}$ from the list of flat connections \cite{Kim:2009wb, Alday:2013rs}. In this paper we are interested in dimension $\Delta=0$ operators, and so we get the condition on $\mathbf{m}$:
\begin{equation}
\label{eqn:mCondition}
0=\frac{1}{4}\sum_{\mathbf{s}}([\textbf{m}\cdot\textbf{s}]_k - [\textbf{m}\cdot\textbf{s}]^2_k/k)-\frac{1}{2}\sum_{i}\left([2m_i]_k-[2m_i]^2_k/k\right).
\end{equation}
We note that the index, as it stands, computes the contribution of $\frac{1}{8}$BPS states since it was defined by picking out a subalgebra spanned by a particular supercharge. In this theory, there exist other protected multiplets that are even shorter. For example, the $\frac{1}{4}$BPS states are annihilated by one more supercharge. The $\frac{1}{4}$BPS contribution to the index can be probed by taking the Macdonald limit~\cite{Gadde:2011uv} of $I_0(\mathbf{m})$, which in this notation is $p\to 0$, keeping $q, t$ fixed. Therefore, only if the exponent vanishes will the index be nonzero. This is the criterion for the allowed topological sectors $\mathbf{m}$ that lead to dimension zero operators. 

To compare to the VW dual SU$(2)_k$ Chern-Simons theory convention, we rescale each $m_i$ by 2 and denote $M_i = 2m_i$. The $M_i$'s are even numbers that correspond to weights of SU$(2)$, with $M_i=2$ corresponding to the positive root of SU$(2)$. The constraint \eqref{eqn:mCondition} now says that $M_3$ must be in the following set:
\begin{equation}
\label{eqn:constraint}
    M_3 \in \{\abs{M_1-M_2}, \abs{M_1-M_2}+2, \abs{M_1-M_2}+4,..., \min(M_1+M_2,2k-M_1-M_2)\}
\end{equation}
which coincides with the SU$(2)_k$ fusion rule. In other words, three weights $a, b, c$ of SU$(2)$ corresponding to the numbers $M_1, M_2, M_3$ satisfy \eqref{eqn:constraint} if an only if the fusion coefficient $N_{abc}$ of SU$(2)_k$ is nonzero (in which case it is invariably $1$). This agrees with the prediction of the VW duality: the allowed states on the dual Chern-Simons side are those that have nonzero $N_{ijk}$, where $N_{ijk}$ is computed from the SU$(k)_2$ fusion rules. By the level-rank duality, this is the same as the SU$(2)_k$ fusion rules. It is crucial that the allowed states must lie on the root lattice. Level rank duality does not preserve the number of states that are not invariant under the outer-automorphism group of the affine Lie algebra. 

Are there any more constraints that the flat connections must satisfy? According to the VW duality mentioned in~\secref{sec:review}, the dimension zero operators of the SU$(2)$ theory on a genus-2 Riemann surface $\Sigma$ and ADE singularity $\mathbb{Z}_k$ are dual to a subset of ground states of level $k$ Chern-Simons theory (those that lie on the root lattice) quantized on the same Riemann surface $\Sigma$. Since the only constraint on the Chern-Simons side is the fusion rule constraint, we expect that the constraint on the flat connections on the $\ClassS$ theory side is the one given by demanding that the Macdonald limit of the index is nonzero. In the next subsection, we show that this conjecture holds for an example of nonabelian $\Gamma$.

As mentioned below \eqref{eqn:casimir1}, the Casimir energies needed to be reduced by an $\mathbf{m}$-independent constant to get dimension $\Delta=0$ operators \cite{Benini:2011nc, Fluder:2017oxm}. We note that in theories with $\NS=1$ SUSY  with only one $R$ charge, it was shown that the supersymmetric Casimir energy is scheme independent, due to the $R$ current symmetry descending uniquely from a supergravity background \cite{Assel:2015nca, Martelli:2015kuk}, and this was generalized in \cite{Bobev:2015kza} to a conjecture about the Casimir energy for theories in arbitrary even dimensions $d$ (compactified on $S^{d-1}$). 


\subsection{Counting holonomy sectors for Dic$_2$}
\label{sec:dic2main}

We now turn to the computation of the Casimir energy, via the superconformal index, in the case where $\Gamma$ is a dicyclic group, and we begin with a brief review of the properties of such groups. A dicyclic group Dic${}_k$ is an extension of the cyclic group of order $2k$ by a group element of order $2$ leading to a non-abelian group with $4k$ elements \cite{nicholson2012introduction}. A set of generators is given by two elements $(r,s)$ satisfying the following relations:
\begin{equation}
\label{eqn:GenDic}
    r^{2k}=1, \quad s^2=r^k, \quad s^{-1}rs=r^{-1}.
\end{equation}
Geometrically, $r$ acts as a rotation and $s$ as an inversion. The case $k=1$ is special, because in this case the only generator of the group is $s$, and it satisfies $s^4=1$. This means that $\text{Dic}_1$ is isomorphic to $\mathbb{Z}_4$, a group we already dealt with in \secref{sec:lens}. In the rest of this section, we compute the single letter index for the first nonabelian example -- $\text{Dic}_2$. We still take the gauge group to be SU$(2)$, so that the flat holonomies are SU$(2)$ representations of $\text{Dic}_2$ up to identifications by SU$(2)$ conjugation. The inequivalent SU$(2)$ representations of $\text{Dic}_k$ for the case of even $k$ can be found in~\cite{Ju:2023umb}. For the specific case $k=2$, the representations for the two generators $r$ and $s$ are listed in the \tabref{tab:RepsDic}.

\begin{table}[ht]
\centering
$\begin{array}{|c|c|c|}
    \hline
    m \ & g(r) & g(s)  \\
    \hline & & \\
    1 & \begin{pmatrix} 1 & 0 \\ 0 & 1 \end{pmatrix} & \begin{pmatrix} 1 & 0 \\ 0 & 1 \end{pmatrix}  \\ & & \\
    \hline & & \\
    2 & \begin{pmatrix} 1 & 0 \\ 0 & 1 \end{pmatrix} & \begin{pmatrix} -1 & 0 \\ 0 & -1 \end{pmatrix} \\ & & \\
    \hline  & & \\
    3 & \begin{pmatrix} -1 & 0 \\ 0 & -1 \end{pmatrix} & \begin{pmatrix} 1 & 0 \\ 0 & 1 \end{pmatrix} \\  & & \\
    \hline  & & \\
    4 & \begin{pmatrix} -1 & 0 \\ 0 & -1 \end{pmatrix} & \begin{pmatrix} -1 & 0 \\ 0 & -1 \end{pmatrix} \\  & & \\
    \hline  & & \\
    5 & \begin{pmatrix} e^{\pi i/2} & 0 \\ 0 & e^{-\pi i/2} \end{pmatrix} & \begin{pmatrix} 0 & 1 \\ -1 & 0 \end{pmatrix} \\ & & \\
    \hline
\end{array}$
\caption{Inequivalent embeddings of Dic$_2$ in SU$(2)$, $g(r)$ and $g(s)$ are the SU$(2)$ matrices that represent the generators $r,s$.}
\label{tab:RepsDic}
\end{table}

The holonomy triplet $abc$ associated to the Trinion (see \figref{fig:trinionSketch}, with $m_1, m_2, m_3$ replaced by labels $a,b,c$) is built out of the individual holonomies listed in the table. That is, for each of $a,b,c$ we choose one of the $5$ representations in \tabref{tab:RepsDic}, and we will write $a,b,c=1,\dots, 5$, referring to this (arbitrary) order in the table. Below, we outline the salient points of the computation of the index for the holonomy triplets, leaving the details for  \appref{app:dic2}. 

The contribution to the index can be broken down into two contributions, one from the vector multiplet and one from the half-hypermultiplet. We first focus on the contribution from a vector multiplet, whose transformation [under $\pi_1(S^3/\Gamma)=\Gamma$] is determined by one of the three holonomies $a,b,$ or $c$ in the holonomy triplet $abc$. Thus, we start by computing $I^V_0(m)$ for all $m\in\{1,2,3,4,5\}$, where later we will replace $m$ with one of $a,b,$ or $c$.

Note that the holonomies 1-4 are abelian, and they act trivially on the vector multiplet, since its fields are all in the adjoint representation. The single letter index can be computed in the same way as for $\Gamma=\mathbb{Z}_k$. The new ingredient here is that, for each holonomy sector $g_m(r)$ and $g_m(s)$, a single letter operator must satisfy two conditions in order to contribute to the index, one from the $r$ generator and one from the $s$ generator. Let $\Phi$ denote a generic BPS field with $j_1=0$, transforming in the adjoint representation. The constraints on the descendants $\partial_{+\dot{+}}^{n_1}\partial_{-\dot{+}}^{n_2}\Phi$ are
\begin{align}
    \label{eq:constraint1}
    r \partial_{+\dot{+}}^{n_1}\partial_{-\dot{+}}^{n_2}\Phi &= \partial_{+\dot{+}}^{n_1}\partial_{-\dot{+}}^{n_2} g_m(r)\Phi g_m(r)^{-1}, \\
    \label{eq:constraint2}
    s \partial_{+\dot{+}}^{n_1}\partial_{-\dot{+}}^{n_2}\Phi &= \partial_{+\dot{+}}^{n_1}\partial_{-\dot{+}}^{n_2} g_m(s)\Phi g_m(s)^{-1},
\end{align}
where the action of $r$ and $s$ on the operator is given by
\begin{align}
    \label{eq:rconstraint}
    r \partial_{+\dot{+}}^{n_1}\partial_{-\dot{+}}^{n_2}\Phi &=  e^{\pi i(n_1-n_2)/2}\partial_{+\dot{+}}^{n_1}\partial_{-\dot{+}}^{n_1}\Phi, \\
    \label{eq:sconstraint}
    s \partial_{+\dot{+}}^{n_1}\partial_{-\dot{+}}^{n_2}\Phi &= e^{\pi in_2}\partial_{+\dot{+}}^{n_2}\partial_{-\dot{+}}^{n_1}\Phi.
\end{align}
Equation~\eqref{eq:sconstraint} implies that, generically, a descendant $\partial_{+\dot{+}}^{n_1}\partial_{-\dot{+}}^{n_2}\Phi$ by itself cannot satisfy the constraint, so one must consider the most general linear combination of single letter descendants (with $j_1=0$): 
\begin{equation}
\Psi_{(j_1=0)} \equiv \sum_{n_1,n_2}C_{n_1n_2}  \partial_{+\dot{+}}^{n_1}\partial_{-\dot{+}}^{n_2}  \Phi,
\end{equation}
where $C_{n_1n_2}$ are constants. We demand that the operator satisfies the constraints
\begin{align}
    \label{eq:rconstraint2}
    r \Psi_{(j_1=0)} &=  g_m(r)\Psi_{(j_1=0)}g_m(r)^{-1},\\
    \label{eq:sconstraint2}
    s \Psi_{(j_1=0)} &= g_m(s)\Psi_{(j_1=0)}g_m(s)^{-1},
\end{align}
from which we solve for the allowed $n_1$ and $n_2$. This will tell us what letters contribute in the single letter index. Since we are only interested in the Casimir energy, we can set the fugacities $p,q$ to $1$ and only sum over contributions from $t$ for simplicity:
\begin{equation}
    \sum t^{2(\Delta+j_2)}.    
\end{equation}

The nonabelian holonomy number $5$ in \tabref{tab:RepsDic} introduces a new ingredient that is different from the lens space case. Unlike the action of the abelian holonomies 1-4 on the BPS field $\Phi$, which is trivial, the action of holonomy 5 can mix different adjoint components of $\Phi=(\Phi_0,\Phi_1,\Phi_{-1})$ (where indices for field components refer to an expansion in a Cartan-Weyl basis):
\begin{align}
    \label{eq:vectormultipletholonomyaction1}
    g_5(r)(\Phi_0, \Phi_1, \Phi_{-1})g_5^{-1}(r) &= (\Phi_0, -\Phi_1, -\Phi_{-1}), \\
    \label{eq:vectormultipletholonomyaction2}
    g_5(s)(\Phi_0, \Phi_1, \Phi_{-1})g_5^{-1}(s) &= (-\Phi_0, -\Phi_{-1}, -\Phi_{1}). 
\end{align}

In particular, note that $g_5(s)$ exchanges $\Phi_1$ with $\Phi_{-1}$. This means that, when we look for the solutions to the constraint imposed by the last holonomy, we need to take linear combinations of the components of $\Phi$ as well:
\begin{align}
\Psi'_{(j_1=0)} &\equiv \sum_{n_1,n_2,p} C_{n_1n_2 p}  \partial_{+\dot{+}}^{n_1}\partial_{-\dot{+}}^{n_2} \Phi_p,
\end{align}
where $p\in \{-1,0,1\}$. The operator $\Psi'$ satisfies the same two constraints as in~\eqref{eq:rconstraint2} and~\eqref{eq:sconstraint2}. This concludes the discussion for components of the vector multiplet with $j_1=0$.

Now, we move on to components of the vector multiplet with spin $j_1=\pm 1/2$. They transform non-trivially under the generator $s$ of Dic${}_2$. We denote these fields by $\lambda_{\mu}$, with $\mu=\pm$ corresponding to the value of the $j_1$ charge.
We begin with the abelian holonomies 1-4 from \tabref{tab:RepsDic}. The most general single letter index operator for holonomies 1-4 is
\begin{align}
\Psi_{(j_1=\pm 1/2)} &\equiv \sum_{n_1,n_2,\mu} C_{n_1n_2\mu}\partial_{+\dot{+}}^{n_1}\partial_{-\dot{+}}^{n_2}\lambda_\mu.
\end{align}

For the nonabelian holonomy $5$ we need to add another index $p\in\{-1,0,1\}$ (in analogous way as before) to capture the nontrivial transformation of the different components of the adjoint representation of the gauge group SU$(2)$:
\begin{align}
\Psi'_{(j_1=\pm 1/2)} &\equiv \sum_{n_1,n_2,\mu,p} C_{n_1n_2\mu p}\partial_{+\dot{+}}^{n_1}\partial_{-\dot{+}}^{n_2}\lambda_{\mu p}.
\end{align}
These letters must also satisfy constraints similar to~\eqref{eq:rconstraint2} and~\eqref{eq:sconstraint2}.

Having discussed the vector multiplet, we now turn to the half-hypermultiplet, which transforms in the product representation of the three holonomies of the triplet $\mathbf{m}=(a,b,c)$, where $a,b,c\in \{1,2,3,4,5\}$. All half-hypermultiplet BPS fields have $j_1=0$, so a generic single letter operator constructed out of a BPS field $\Phi_{ijk}$ (where $i,j,k\in \{-1,1\}$ denote the indices for the SU$(2)^3$ trifundamental representation) is
\begin{equation}
    \Psi \equiv \sum_{n_1,n_2,i,j,k} C_{n_1n_2ijk}   \partial_{+\dot{+}}^{n_1}\partial_{-\dot{+}}^{n_2} \Phi_{ijk}.
\end{equation}
It must satisfy two constraints similar to what the vector multiplet operator satisfies in equations~\eqref{eq:rconstraint2} and~\eqref{eq:sconstraint2}, except that the holonomies are now in the fundamental representation. For the holonomy triplet $abc$, the constraints are
\begin{align}
    r \Psi  &=  g_a(r)g_b(r)g_c(r)\Psi, \\
    s \Psi &= g_a(s)g_b(s)g_c(s)\Psi.
\end{align}
The holonomies $g_a(r), g_a(s)$ act only on the base field components $\Phi_{ijk}$ in $\Psi$ and do not act on the derivatives $\partial_{\pm \dot{+}}$. The base field $\Phi_{ijk}$ transforms under the holonomy triplet $abc$ as
\begin{equation}
    \label{eq:hypermultipletholonomyaction}   
     g_a(r)g_b(r)g_c(r) \Phi_{ijk} = \sum_{i',j',k'} (g_a(r))_{ii'} (g_b(r))_{jj'} (g_c(r))_{kk'} \Phi_{i'j'k'}
\end{equation}
and similarly for $s$. This concludes the setup of the problem of computing the single letter index on $S^3/\text{Dic}_2$. The details of the computation are left for \appref{app:dic2}, and in the following we present the final results.


\subsubsection{Trinion contribution to the Casimir energy on $S^3/\text{Dic}_2$}
\label{subsubsec:dic2list}
After we decompose a Riemann surface $\Sigma$ into trinion vertices connected by edges to construct a Lagrangian for the $\ClassS$ theory, as in \figref{fig:2}, and after selecting an SU$(2)$ flat connection from the list $1,\dots,5$ for each edge (i.e., for each vector multiplet), the total Casimir energy for the particular configuration of flat connections is the sum of contributions from edges and vertices, computed in \appref{app:dic2}. It is convenient to split the contribution of each edge into two halves, and add the halves to the contributions of the two vertices attached to the edge. After that, to compute the Casimir energy of configuration of holonomies, we just need to add the contribution of the vertices. Each vertex is labeled by a triplet $(ijk)$, with $1\le i,j,k\le 5$ corresponding to the SU$(2)$ flat connections on $S^3/\text{Dic}_2$ of the edges attached to the vertex. From the calculation of \appref{app:dic2} we get this modified contribution of each vertex, which we list below. In the list, we group the holonomy triplets into four distinct sets that each have the same supersymmetric Casimir energy:

\begin{itemize}
    \item 
There are 28 triplets that have the same energy as the trivial holonomy triplet (111). These are: (122), (133), (144), (234), (155), (255), (355), (455). In total there are 28 combinations, since we need to count different orderings [e.g., $(155)$, $(515)$, and $(551)$] as distinct combinations. These triplets have the lowest positive supersymmetric Casimir energy $E_0 = \frac{35}{108}$.

\item
The holonomy triplets of the form $(ab5)$, $a,b\in\{1,2,3,4\}$ and their associated permutations all have the supersymmetric Casimir energy $E_1= \frac{143}{108}$. We note that $E_1=E_0+1$.

\item
The holonomy triplets of the form (112), (113), (114), (233), (244), (222), (134), (224), (334), (223), (344), (333), (444), (123), (124) and their associated permutations have the same supersymmetric Casimir energy $E_2=\frac{251}{108}=E_0+2$.

\item
Finally, the holonomy triplet (555) has a negative supersymmetric Casimir energy, $E_3=-\frac{19}{108}=E_0-\frac{1}{2}$. 

\end{itemize}


\subsubsection{Combining the vertex contributions for $\Sigma$}
\label{subsubsec:dic2combine}

We decompose the Riemann surface $\Sigma$ into trinions, label each edge by a flat connection $1,\dots,5$, and add the contributions of the vertices according to the list at the end of \secref{subsubsec:dic2list}. For example, if the edges in \figref{fig:2} are $a=1$ and $b=c=2$ we need twice the contribution of the (122) triplet, and the Casimir energy is $2E_0$, while if $a=b=1$ and $c=2$ we get $2E_0+4$, and if $a=b=c=5$, we get $2E_0-1$.

In general, for a Riemann surface of genus $g>1$, if all the vertices ($2g-2$ in number) are labeled by triplets from the first item of the list in \secref{subsubsec:dic2list}, the resulting Casimir energy will be $E=(2g-2)E_0$. This is also the Casimir energy for the trivial flat connection (i.e., all edges labeled by $1$), and we expect these states to correspond to dimension $\Delta=0$ operators on $\mathbb{R}^4/\Gamma$, where $(2g-2)E_0$ is accounted for by the conformal anomaly of the state-operator transformation. We will argue in \secref{subsubsec:dic2CS} that such states precisely agree with the conjectured Chern-Simons dual system.

For genus $g\le 3$, all other combinations of labels on the edges lead to states with Casimir energy $E>(2g-2)E_0$, except when all edges are labeled by $5$, in which case the energy is $E=(2g-2)E_0+1-g$, which we expect to lead to $\Delta=1-g<0$. This does not necessarily contradict unitarity, as we will discuss in \secref{subsubsec:negativeDelta}, but these states are not captured by the proposed Chern-Simons system.

For $g\ge 4$, we can also combine vertices with Casimir energy $E_0+2$ and vertices with Casimir energy $E_0-\frac{1}{2}$ to get a total energy of $E=(2g-2)E_0$, just like the trivial flat connection. These states also do not fit with the proposed Chern-Simons dual, as we explain in \secref{subsubsec:dic2CS}. We expect that under the action of the mapping class group of $\Sigma$, those states do not mix with those that are constructed purely from combinations of the triplets from the first item of the list in \secref{subsubsec:dic2list}:
$$
(111), (122), (133), (144), (234), (155), (255), (355), (455).
$$


\subsubsection{Comparison with $SO(8)_2$ Chern-Simons theory}
\label{subsubsec:dic2CS}
The Chern-Simons theory to which we need to compare the results of \secref{subsubsec:dic2list}, under the VW duality, has an SO$(8)$ gauge group (dual to Dic$_2$ as given by the McKay correspondence in \tabref{tab:ADE}) and level $2$ [since our initial gauge group is SU$(2)$]. The 5 possible holonomies labeled by $a=1,\dots,5$ correspond to points lying on the lattice \eqref{eqn:statelattice2}, which in terms of SO$(8)$ Dynkin labels (i.e., the coefficients in an expansion in the fundamental weights) are given by:
\begin{align}
    (0,0,0,0), \; (2,0,0,0), \; (0,0,2,0), \; (0,0,0,2), \; (0,1,0,0).
\end{align}
With this order, the following holds: a triplet appears in the list of states with Casimir energy $E_0$ precisely if the corresponding fusion coefficient $N_{abc}$ of SO$(8)_2$ is nonzero.

We recall that in the case of $S^3/\mathbb{Z}_k$, we found that $\frac{1}{4}$BPS states which have the same supersymmetric Casimir energy as the trivial holonomy triplet $(abc)$, are dual to Chern Simons ground states on a $3$-punctured sphere with Wilson loops inserted at the punctures in representations dual to $a$, $b$ and $c$. For $S^3/\text{Dic}_2$, we now find the same observation, since holonomy triplets with supersymmetric Casimir energy equal to $E_0$ satisfy the SO$(8)_2$ fusion rule.
We can therefore match the dimension of states with Casimir energy $2E_0$ on $S^3/\text{Dic}_2$ for $\ClassS$ theory of a genus $g=2$ Riemann surface $\Sigma$. The surface $\Sigma$ is obtained by gluing two 3-punctured spheres; the energy doubles; and we count $28$ states, which is precisely the dimension of the SO$(8)_2$ Chern-Simons theory on a closed Riemann surface of genus 2, as can be computed by summing over the fusion coefficients $N_{abc}N^{abc}$. The fusion coefficients $N_{abc}$ for this Chern-Simons theory are nonzero precisely when the index $abc$ is one of the 28 states and vanishes otherwise. This gives strong support to the duality we proposed, and we conjecture that this result generalizes to other D- and E-singularities and for all SU$(N)$ gauge groups of SCFT.

We expect states with energy $2E_0$ to correspond to operators with dimension $\Delta=0$, and we will not discuss operators with dimension $\Delta>0$ in this paper.
We leave the question of the action of the mapping class group of $\Sigma$ on the other states discussed in \secref{subsubsec:dic2combine}, including those with $\Delta<0$, as an open problem.


\subsubsection{A note about $\Delta<0$ and unitarity}
\label{subsubsec:negativeDelta}

Let us explain why $\Delta<0$ becomes possible when $\Gamma$ is nonabelian (without manifestly contradicting unitarity).
In general, BPS states that contribute to the Macdonald limit of the index satisfy
\begin{align}\label{eqn:BPSCondMacd}
    \Delta+2j_1-2R-r=0,
\end{align}
in addition to the general BPS condition $\delta=0=\Delta - 2j_2 - 2R + r$. Note that in \eqref{eqn:BPSCondMacd} the sign of $r$ is opposite to the one in $\delta$.
Since $\Gamma\subset\text{SU}(2)_1$,  the SU$(2)_2$ and SU$(2)_R$ groups are symmetries of the system on $S^3/\Gamma$, and therefore states fall into multiplets of SU$(2)_2$ and SU$(2)_R$ but not of SU$(2)_1$. 
The operators that contribute to the index are those components of the multiplet with highest $j_2$ and $R$. (This is because the superconformal algebra requires $\delta\ge 0$ for all states and if there was a state with higher $j_2$ or $R$ in the same multiplet as a state with $\delta=0$ it would have negative $\delta$.) Since the highest $j_2$ and $R$ of an $\text{SU}(2)_2\times\text{SU}(2)_R$ multiplet are nonnegative, we arrive at the bounds
\begin{align}\label{eqn:Deltarj1}
    \Delta+r\geq 0, \quad \Delta+2j_1-r\geq 0,
\end{align}
that have to be satisfied by any operator that contributes to the index in the Macdonald limit.
If we could show that for any operator that contributes to the index and has dimension $\Delta$ and charges $r$ and $j_1$, there is a corresponding operator with the same dimension but with charges $r$ and $-j_1$, then we would also have $\Delta-2j_1-r\geq 0$, and summing we would get the counterfactual
$$
4\Delta=2(\Delta+r)+(\Delta+2j_1-r)+(\Delta-2j_1-r)\stackrel{?}{\ge} 0,
$$
where we use the $\stackrel{?}{\ge}$ notation to indicate that this inequality does not necessarily hold.
In the case of $\Gamma=\mathbb{Z}_n$ there is a discrete rotation $\Omega$ element in $\text{Spin}(4)$ whose adjoint action fixes $\Gamma\subset\text{Spin}(4)$ and thus $\Omega$ is preserved by the orbifold projection and allows us to argue indeed that the sought after states with charges $-j_1$ and same $r$, and $\Delta$ do exist. Explicitly, $\Omega$'s action on a state with quantum numbers $R,r,j_1,j_2$ is: $\Omega|R,r,j_1,j_2\rangle=(-1)^R|R,r,-j_1,j_2\rangle$, which implies that for $\Gamma=\mathbb{Z}_n$ we have the bound $\Delta\geq 0$.

However, the above argument fails for $\Gamma=\text{Dic}_2$ because it is nonabelian \eqref{eqn:GenDic} and there is no conserved quantum number $j_1$ anymore. The BPS bounds are less restrictive, and one could potentially find states with $\Delta<0$. 

Operators with negative conformal dimensions also naturally arise in a 2D orbifold scenario. Consider a 2D CFT with central charge $c$ on a cylinder parameterized by $0\le\sigma<2\pi$ and $-\infty<\tau<\infty$ with periodic complex coordinate $z=\sigma+i\tau$. (We will assume that the left moving and right moving CFTs have the same central charge, for simplicity.) The Casimir energy is $-c/24$. Now define a coordinate on $\mathbb{C}$ by $w=\exp(i z/k)$ for some integer $k$. The coordinate $w$ is multivalued and descends to a single-valued coordinate on $\mathbb{C}/\mathbb{Z}_k$, where the orbifold identification is $w\sim \exp(2\pi i/k)w$.
The state-operator correspondence maps the ground state on the cylinder to an operator of dimension
$$
\Delta = -\frac{c}{24}+\frac{c}{24k^2}
$$
computed via the Schwarzian derivative $\{e^{-i z/k},z\}$. For $k=1$ the ground state maps to the operator $1$ of dimension $\Delta=0$, but for $k>1$ we find $\Delta<0$.


\section{Counting ground states for genus two with $G=\textmd{SU}(N)$ and $\Gamma=\mathbb{Z}_2$}
\label{sec:counting}
In section~\secref{sec:review}, we started with an example of counting the ground states of the SU$(2)$ theory on a genus-2 Riemann surface with an $A_1$ singularity ($\Gamma=\mathbb{Z}_2$). The $\ClassS$ theory with SU$(2)$ gauge group enjoys an explicit known Lagrangian description in terms of trifundamental hypermultiplets, which was used to arrive at the constraint in equation~\eqref{eqn:constraint}, which in turn led to a verification of the conjectured VW duality for that case, at the end of section~\secref{sec:index}. In this section, we count the ground states for gauge group SU$(N)$ with $N>2$ theory on a genus-2 Riemann surface (see \figref{fig:2}) times an ADE singularity $\mathbb{R}^4/\mathbb{Z}_2$. To the best of our knowledge, no Lagrangian description is currently known for the trinion theories with $N>2$, although partial results are available: for example, the $N=3$ trinion theory can be expressed \cite{Gaiotto:2009we} in terms of the Minahan-Nemeschansky $E_6$ SCFT \cite{Minahan:1996fg}, whose partially gauged form can be expressed as a limit of a strongly coupled Lagrangian SCFT \cite{Argyres:2007cn}. 

The flat connections can be represented by $N\times N$ diagonal matrices $(\pm1)$ eigenvalues. Denoting such a matrix by ${\mathbf{M}}$, the number of $(-1)$'s in ${\mathbf{M}}$ must be even in order to satisfy $\det\mathbf{M}=1$, as ${\mathbf{M}}$ represents the generator of $\mathbb{Z}_2$ in a group homomorphism $\mathbb{Z}_2\rightarrow\text{SU}(N)$. Therefore, up to identifications by the Weyl group $S_N$ of SU$(N)$, there are in total $\lfloor N/2\rfloor+1$ flat connections. Let $a, b, c$ denote the three SU$(N)$ flat connections in the theory of \figref{fig:2} that is built from two $T_N$ theories coupled with $\text{SU}(N)^3$ gauge multiplets. Let $g_a, g_b, g_c \in \text{SU}(N)$ be representations of these flat connections as diagonal matrices with $\pm1$ eigenvalues. Let $w\in S_N$ be a permutation that acts on ${\mathbf{M}}$ as a Weyl group element $\mathbf{M}\mapsto w(\mathbf{M})$, i.e., by permuting the elements on the diagonal of $\mathbf{M}$.
We claim that a basis of ground states of the double-$T_N$ theory can be labeled as
\begin{equation}
    (g_a, g_b , g_c),
\end{equation}
where $g_a, g_b, g_c$ satisfy that their product for at least one rearrangement of the eigenvalues is the identity matrix:
\begin{equation}
\label{eqn:abc}
    w_1(g_a) w_2(g_b) w_3(g_c) = \textbf{I},\qquad\text{for some $w_1, w_2,w_3\in S_N$.}
\end{equation}

According to the conjecture of VW duality, the ground states of the $\ClassS$ theory must match triplets of representations of SU$(2)_N$ Chern Simons theory for which the fusion coefficient is nonzero. We claim that equation~\eqref{eqn:abc} encodes exactly the information of the fusion rules. A proof sketch is as follows.

We denote a flat connection by $g_k$ where $k$ is the number of $(-1)$'s along the diagonal. By assumption, $0\leq k\leq N$ and $k$ is even. We ask the question: if we multiply two such matrices together, say $g_l$ and $g_m$, what can the solution be modulo Weyl symmetry? We first notice that the least number of $(-1)$'s we can get in the resulting product is by overlapping as many $(-1)$'s from $g_l$ as possible with $(-1)$'s from $g_m$. Without loss of generality, let us assume $l\ge m$. It is not hard to see that the least number of $(-1)$'s is 
\[
l-m,
\]
which means that $g_{l-m}$ is a possible value of the product $w_1(g_l) w_2(g_m)$ (for some $w_1, w_2\in S_N$). For example, setting $N=5$, $l=4$, and $m=2$, we have, for the least number of $-1$s:
\[
w_1(g_4)w_2(g_2) =
\begin{pmatrix}
-1 & 0 & 0 & 0 & 0 \\
0 & -1 & 0 & 0 & 0 \\
0 & 0 & -1 & 0 & 0 \\
0 & 0 & 0 & -1 & 0 \\
0 & 0 & 0 & 0 & 1 \\
\end{pmatrix}
\begin{pmatrix}
-1 & 0 & 0 & 0 & 0 \\
0 & -1 & 0 & 0 & 0 \\
0 & 0 & 1 & 0 & 0 \\
0 & 0 & 0 & 1 & 0 \\
0 & 0 & 0 & 0 & 1 \\
\end{pmatrix}
=
\begin{pmatrix}
1 & 0 & 0 & 0 & 0 \\
0 & 1 & 0 & 0 & 0 \\
0 & 0 & -1 & 0 & 0 \\
0 & 0 & 0 & -1 & 0 \\
0 & 0 & 0 & 0 & 1 \\
\end{pmatrix}
=w_3(g_2),
\]
for a suitable $w_3\in S_N$.

We can compute the maximum number of $(-1)$'s that we can get by multiplying $w_1(g_l)$ and $w_2(g_m)$ as follows. Starting with a configuration that yields the least number of $(-1)$'s in the product, we can move one $-1$ in $g_m$ to a position where the corresponding eigenvalue in $g_l$ is a $(+1)$. This operation increases the number of $(-1)$'s in $w_1(g_l)w_2(g_m)$ by 2, yielding $g_{l-m+2}$, up to a Weyl permutation. In our example, this procedure is illustrated by shifting the second $(-1)$ in $g_2$ to the last position on the diagonal:
\[
w_1(g_4)w_2'(g_2) =
\begin{pmatrix}
-1 & 0 & 0 & 0 & 0 \\
0 & -1 & 0 & 0 & 0 \\
0 & 0 & -1 & 0 & 0 \\
0 & 0 & 0 & -1 & 0 \\
0 & 0 & 0 & 0 & 1 \\
\end{pmatrix}
\begin{pmatrix}
-1 & 0 & 0 & 0 & 0 \\
0 & 1 & 0 & 0 & 0 \\
0 & 0 & 1 & 0 & 0 \\
0 & 0 & 0 & 1 & 0 \\
0 & 0 & 0 & 0 & -1 \\
\end{pmatrix}
=
\begin{pmatrix}
1 & 0 & 0 & 0 & 0 \\
0 & -1 & 0 & 0 & 0 \\
0 & 0 & -1 & 0 & 0 \\
0 & 0 & 0 & -1 & 0 \\
0 & 0 & 0 & 0 & -1 \\
\end{pmatrix}
=w_3'(g_4).
\]

We can keep performing this process until we reach a product that has the maximal number of $(-1)$'s. The matrix $g_l$ has $N-l$ elements $(+1)$. If we can manage to shift all $(-1)$'s in $g_m$ to those $N-l$ slots, the resulting product will contain $l+m$ elements $(-1)$'s. 
There are now two cases to consider. If $m+l\leq N$, we can accomplish that, and the last matrix in the sequence of operations will be $g_{m+l}$. However, if $m+l>N$, there will be some inevitable overlaps of $(-1)$'s for the two matrices $w_1(g_l)$ and $w_2(g_m)$. The minimal number of overlaps is $l+m-N$, and the corresponding product $w_1(g_l)w_2(g_4)$ has $N-(l+m-N)=2N-(l+m)$ elements $(-1)$, giving the matrix $g_{2N-(l+m)}$.
The final matrix in this process of generating $(-1)$'s therefore contains
\[
\min(m+l, 2N-m-l)
\]
$(-1)$'s. The multiplication rule that we have thus derived is
\[
w_1(g_l)w_2(g_m) \longrightarrow\{ g_{\abs{l-m}}, g_{\abs{l-m}+2},..., g_{\min(m+l, 2N-m-l)}\},
\]
which is exactly the SU$(2)_N$ fusion rule~\eqref{eqn:constraint}. This shows that the ground states of the non-Lagrangian SU$(N)$ theory on $\Sigma$ and $\mathbb{Z}_2$ singularity can be derived from the rule \eqref{eqn:abc}.
However, we do not have a first principles derivation of \eqref{eqn:abc} on the SCFT side. We merely observe that assuming \eqref{eqn:abc}, the number of ground states match.
 
\section{Generalization and conclusion}
\label{sec:generalization}

In this work, we put forward a conjecture about the action of the duality group on dimension $\Delta=0$ operators of a $\ClassS$ theory formulated on $\mathbb{R}^4/\Gamma$, where the operators are at the singularity. The conjecture, motivated by Vafa and Witten's results for $\NS=4$ SYM \cite{Vafa:1994tf} is that it is equivalent to the action of the mapping class group of the Riemann surface $\Sigma$ (underlying Gaiotto's construction of the $\ClassS$ theory) on the Hilbert space of Chern-Simons theory on $\Sigma$ with gauge group determined by $\Gamma$, through the McKay correspondence, and the level determined by the gauge group of the $\ClassS$ theory.
We provided some modest supporting evidence by matching the number of ground states of the relevant Chern-Simons theory with the number of $\Delta=0$ operators at the singularity, for $\Gamma=\mathbb{Z}_k$ and $\Gamma=\text{Dic}_2$, corresponding to Chern-Simons gauge groups SU$(k)$ and SO$(8)$. The counting for the abelian $\mathbb{Z}_k$ proceeded in the same way as previously computed in the literature in the context of the 4d/2d and 3d/3d correspondence \cite{Benini:2011nc, Alday:2013rs, Gukov:2015sna}, with the difference that our Chern-Simons theory parameters are connected by level-rank duality to a {\it compact} form of the noncompact Chern-Simons theory that appears in the 3d/3d correspondence, but the calculation is otherwise identical. It relies on the nontrivial Casimir energies that arise for some combinations of flat connections when the $\ClassS$ theory is formulated on $S^3/\Gamma$. (For $\NS=4$ SYM, such a Casimir energy does not produce constraints on holonomies, as demonstrated explicitly in \cite{Ju:2023ssy}.) A novel aspect of our results, to the best of our knowledge, is the extension of the calculation to a nonabelian $\Gamma=\text{Dic}_2$ where the $\Delta=0$ operators again match with Chern-Simons states, thanks to restrictions due to nontrivial Casimir energies. For the nonabelian $\Gamma$, we also found a state with $\Delta<0$, which is another novelty.
It would be interesting to explore correlation functions of these operators, say, on $T^4/\text{Dic}_2$, where the action of $\text{Dic}_2$ is defined so as to produce two singularities that look locally as $\mathbb{R}^4/\text{Dic}_2$, as well as additional singularities of type $A_1, A_3$.
Extensions to other nonabelian $\Gamma$'s, i.e., corresponding to $E_6, E_7, E_8$, or $\text{Dic}_k$ with $k>2$, will be explored elsewhere. Extensions to $\ClassS$ theories derived from D-type or E-type $(2,0)$-theory are also left for future exploration.

\acknowledgments
We wish to thank Sergey Cherkis, Petr Ho\v{r}ava, Jesus Sanchez Jr, and Yasunori Nomura for illuminating discussions. This work has been supported by the Berkeley Center for Theoretical Physics.

\appendix

\section{Illustration of Casimir Energy computation with index}
\label{app:CasimirExample}
We give an illustration of the Casimir energy formula~\eqref{eqn:CasimirDef} using the example of a free chiral NS fermion on $S^1$. 
The modes are $\psi_{-n-\frac{1}{2}}$, with $n=0,1,2,\dots$, the states are
$$
\psi_{-n_1-\frac{1}{2}}
\psi_{-n_2-\frac{1}{2}}
\cdots
\psi_{-n_r-\frac{1}{2}}
\ket{0}.
$$
Under the state-operator correspondence, this state corresponds to the operator
$$
\partial^{n_1}\psi\partial^{n_2}\psi\cdots\partial^{n_r}\psi.
$$
The partition function is
$$
Z=
e^{-\beta E_0}\prod_{n=0}^\infty\left(1-e^{-(n+\frac{1}{2})\beta}\right),
$$
where $E_0$ is the Casimir energy.
We have
$$
\left(1-e^{-(n+\frac{1}{2})\beta}\right)=\text{PE}\left[-e^{-(n+\frac{1}{2})\beta}\right],
$$
and
$$
Z = e^{-\beta E_0}\text{PE}\left[-\sum_{n=0}^\infty e^{-(n+\frac{1}{2})\beta}\right],
$$
the single particle index is
$$
I=-\sum_{n=0}^\infty e^{-(n+\frac{1}{2})\beta},
$$
and
$$
-\frac{1}{2}\left\lbrack
\frac{\partial I}{\partial\beta}\right\rbrack
_{\beta\rightarrow 0}\longrightarrow
-\frac{1}{2}\sum_{n=0}^\infty (n+\frac{1}{2})
$$
is the (divergent) Casimir energy.
To confirm, we calculate
$$
I=-\frac{e^{-\beta/2}}{1-e^{-\beta}}
=-\frac{1}{\beta}+\frac{\beta}{24}+\cdots,\qquad
-\frac{1}{2}\left\lbrack
\frac{\partial I}{\partial\beta}\right\rbrack
=-\frac{1}{2\beta^2}-\frac{1}{48} + O(\beta^2).
$$
We get the regularized Casimir energy of $-\frac{1}{48}$ after dropping the divergent term $\frac{1}{\beta^2}$ and taking the $\beta\to 0$ limit.

\section{Details of $S^3/\text{Dic}_2$ index calculation}

\label{app:dic2}

In this section, we give a detailed computation of the superconformal single letter index on $S^3/\text{Dic}_2$. The idea behind the computation is discussed in \secref{sec:dic2main}. A quick summary is that, for a BPS operator to contribute to the single letter index, it must satisfy two constraints as in equations~\eqref{eq:constraint1} and \eqref{eq:constraint2}. The contribution of an operator to the index is given by the Boltzmann weight $t^{2(\Delta+j_2)}$, where for each field its charges $(\Delta,j_2)$ are given in \tabref{tab:SingleLetter}. In this appendix, we will explicitly solve for the constraints and sum over the appropriate Boltzmann weights in order to find the index in the case of the $\text{Dic}_2$ group. 
We recall that there are five distinct flat connections (see \tabref{tab:RepsDic}) which we label by $j\in\{1,2,3,4,5\}$. The first four are abelian, with $j=1$ labeling the trivial connection, and $j=5$ is (the only) nonabelian connection. The nontrivial abelian flat connections are $j=2,3,4$. Counting triplets of holonomies, we will find that in total, there are 125 triplets, given by the following groupings:

\begin{itemize}
    \item $(111)$: 1 triplet involving only the trivial holonomy.
    \item $(115)$: 3 triplets involving two trivial holonomies and one nonabelian holonomy. There are 3 holonomy triplets of this type, counting permutations as distinct triplets.
    \item $(11a)$: 9 triplets involving two trivial holonomies and one nontrivial abelian holonomy. Here, $a\in \{2,3,4\}$ represents any of the three nontrivial abelian holonomies. 
    \item $(1ab)$: 18 triplets involving one trivial holonomy and two distinct nontrivial abelian holonomies.
    \item $(1aa)$: 9 triplets involving one trivial holonomy and two identical nontrivial abelian holonomies.
    \item $(j55)$: 12 triplets involving one abelian holonomy ($j\in\{1,2,3,4\}$) and two identical nontrivial nonabelian holonomies.
    \item $(1a5)$: 18 triplets involving one trivial holonomy, one nontrivial abelian holonomy, and one nonabelian holonomy.
    \item $(abc)$: 6 triplets involving three distinct nontrivial abelian holonomies.
    \item $(aab)$: 18 triplets involving two nontrivial abelian holonomies that are identical and one that is distinct from the two.
    \item $(aaa)$: 3 triplets involving three nontrivial abelian holonomies that are identical.
    \item $(ab5)$: 18 triplets involving two nontrivial abelian holonomies that are distinct and one nonabelian holonomy.
    \item $(aa5)$: 9 triplets involving two nontrivial abelian holonomies that are identical and one nonabelian holonomy.
    \item $(555)$: 1 triplet involving three nonabelian holonomies.
\end{itemize}

Although the number of holonomy triplets is large, we will shortly see that many of these triplets will lead to the same single letter index.
We denote a generic BPS field with $j_1=0$ as $\Phi$, and a generic BPS field with $j_1=\pm 1/2$ as $\lambda_{\pm}$. The geometric action of $r$ and $s$ on BPS descendants of fields with $j_1=0$ is discussed in \secref{sec:dic2main} [see equations~\eqref{eq:rconstraint} and~\eqref{eq:sconstraint}, which we repeat here for convenience]:
\begin{align*}
    r  \partial^{n_1}_{+\dot{+}} \partial^{n_2}_{-\dot{+}} \Phi   &= e^{i\pi (n_1-n_2)/2} \partial^{n_1}_{+\dot{+}} \partial^{n_2}_{-\dot{+}} \Phi, \\
    s \partial^{n_1}_{+\dot{+}} \partial^{n_2}_{-\dot{+}} \Phi &= (-1)^{n_2} \partial^{n_2}_{+\dot{+}} \partial^{n_1}_{-\dot{+}} \Phi.
\end{align*}

For fields with $j_1=\pm1/2$ we note that $\lambda_{\pm}$ themselves carry angular momentum and transform into each other as an SU$(2)$ doublet under $s$:
\begin{align*}
    r \partial^{n_1}_{+\dot{+}} \partial^{n_2}_{-\dot{+}}  \lambda_{\mu}  &= e^{ i\pi (n_1-n_2+\mu)/2}  \partial^{n_1}_{+\dot{+}} \partial^{n_2}_{-\dot{+}} \lambda_{\mu},   \\
    s\partial^{n_1}_{+\dot{+}} \partial^{n_2}_{-\dot{+}}\lambda_{\mu} &= (-1)^{(\mu-1)/2} e^{\pi i n_2} \partial^{n_2}_{+\dot{+}} \partial^{n_1}_{-\dot{+}}  \lambda_{-\mu}, 
\end{align*}
where the index $\mu$ corresponds to $\pm1/2$.

In \secref{sec:dic2main} We discussed the effect of the gauge transformation experienced by the fields along a nontrivial $1$-cycle of $S^3/\Gamma$. See equations~\eqref{eq:vectormultipletholonomyaction1} and~\eqref{eq:vectormultipletholonomyaction2} for the vector multiplet and equation~\eqref{eq:hypermultipletholonomyaction} for the hypermultiplet. Therefore, we now have all the ingredients we need to compute the single letter indices. For the contribution of the vector multiplets, we simply compute the contribution of each multiplet separately, and then multiply the results for the three vector multiplets corresponding to the three $SU(2)$ gauge group factors. The contribution of each vector multiplet, separately, depends only on one of the holonomies (labeled by $j=1,\dots,5$) of the holonomy triplet, and we begin by computing that contribution.


\subsection{Contribution of a single vector multiplet $I^{vec}$}

First, we note that the vector multiplet contributions for holonomies $1-4$ are identical,
\[
I^{vec}_1 = I^{vec}_2 =I^{vec}_3=I^{vec}_4,
\]
because when acting on fields in the adjoint representation of the gauge group $SU(2)$, the representations of $\text{Dic}_2$ defined by each of the holonomies $1-4$ are all trivial, and there is no distinction between the four holonomies $1-4$ (see \tabref{tab:RepsDic}).

 We first compute the contribution from the BPS fields $\Phi$ that have spin $j_1=0$. As discussed in \secref{sec:dic2main}, we need to compute the dimension of the vector space of linear combinations of the descendants, of the form
\begin{equation}\label{eqn:Sumcn1n2Phi}
\sum c^{n_1n_2}\partial^{n_1}_{+\dot{+}} \partial^{n_2}_{-\dot{+}} \Phi,
\end{equation}
with constant coefficients $c^{n_1 n_2}$, that satisfy two constraints which follow from the action of the generators $r,s$ of the Dicyclic group:
\begin{align*}
    \sum c^{n_1n_2} \partial^{n_1}_{+\dot{+}} \partial^{n_2}_{-\dot{+}} \Phi&=\sum e^{i\pi /2 (n_1-n_2)} c^{n_1n_2} \partial^{n_1}_{+\dot{+}} \partial^{n_2}_{-\dot{+}} \Phi, \\
    \sum c^{n_1n_2} \partial^{n_1}_{+\dot{+}} \partial^{n_2}_{-\dot{+}} \Phi&=\sum e^{i\pi n_2} c^{n_1n_2} \partial^{n_2}_{+\dot{+}} \partial^{n_1}_{-\dot{+}} \Phi.
\end{align*}
The first equation implies that $c^{n_1n_2}$ must vanish unless
\[
n_1 - n_2 \equiv 0 \pmod{4},
\]
and the second constraint shows that $c^{n_1n_2}$ with $n_1<n_2$ is determined by $c^{n_1 n_2}$ with $n_1>n_2$ and as for the diagonal terms, $c^{n_1 n_1}$ vanish unless $n_1$ is even. 

Using \tabref{tab:SingleLetter}, we find that $\partial^{n_2}_{+\dot{+}} \partial^{n_1}_{-\dot{+}}$ contributes $t^{3n_1+3n_2}y^{n_2-n_1}$ to the single letter index. Let $I_\Phi(t)$ be the contribution of the (as yet unspecified) field $\Phi$ to the single-letter index. Since $\Phi$ is assumed to be in the adjoint representation of $SU(2)$, which is $3$-dimensional, we include an extra factor of $3$ and find the contribution of words of the form \eqref{eqn:Sumcn1n2Phi} to the single-letter index to be, after setting $y=1$,
\begin{equation}\label{eqn:ContribFSumcn1n2Phi}
F=3\left\{\sum_{\tiny
\begin{array}{c}n_1>n_2\ge0\\
n_1\equiv n_2\!\!\!\!\pmod{4}\\
\end{array}} t^{3n_1+3n_2}
+\sum_{\tiny n_1\equiv 0\!\!\!\!\pmod{2}}
t^{6n_1}\right\}I_\Phi(t)
=\frac{3(1-t^6+t^{12})}{(1-t^6)(1-t^{12})}I_\Phi(t)\,.
\end{equation}
Referring to \tabref{tab:SingleLetter}, the possible fields to substitute for $\Phi$ are those with $j_1=0$, which are $\phi$, $\overline{\lambda}_{1\dot{+}}$, and $\overline{F}_{\dot{+}\dot{+}}$.  Setting the fugacity $y=1$, we find $I_\Phi(t)=t^2-t^4+t^6$, and therefore, the contribution of the spin projection $j_1 = 0$  fields in the vector multiplet to the single letter index for holonomies $1-4$ is given by
\begin{equation}
\label{eq:vector1spin0}
\frac{3(1-t^6+t^{12})(t^2-t^4+t^6)}{(1-t^6)(1-t^{12})}\,.
\end{equation}
For the case of fields $\lambda_\mu$ ($\mu=\pm 1$) with spin projections $j_1=\mu/2=\pm 1/2$, a generic operator now has the following form:
\[
\sum c^{\mu n_1n_2} \partial^{n_1}_{+\dot{+}} \partial^{n_2}_{-\dot{+}}\lambda_{\mu}. 
\]
Imposing the algebraic constraints, we obtain
\begin{align*}
    \sum c^{\mu n_1n_2}\partial^{n_1}_{+\dot{+}} \partial^{n_2}_{-\dot{+}} \lambda_{\mu}&=\sum e^{i\pi(n_1-n_2 +\mu)/2}c^{\mu n_1n_2}\partial^{n_1}_{+\dot{+}} \partial^{n_2}_{-\dot{+}} \lambda_{\mu},  \\
    \sum c^{\mu n_1n_2} \partial^{n_1}_{+\dot{+}} \partial^{n_2}_{-\dot{+}} \lambda_{\mu} &=\sum (-1)^{n_2+(\mu-1)/2}c^{\mu n_1n_2} \partial^{n_2}_{+\dot{+}} \partial^{n_1}_{-\dot{+}} \lambda_{-\mu}.
\end{align*}
The first constraint shows that when $\mu=1$ we have
\[
n_1-n_2 \equiv 3\pmod{4},
\]
and the second constraint shows that The $\mu=-1$ coefficients $c^{-1, n_1n_2}$ are completely determined from the $\mu=1$ coefficients $c^{1,n_1n_2}$.
From \tabref{tab:SingleLetter} we find that the fields $\lambda$ carry a $t^3$ factor, and therefore, the contribution of the $\lambda_{\pm}$ fields to the index is 
\begin{equation}
\label{eq:vector1spin1}
-3t^3\sum_{\tiny n_1-n_2\equiv 3\!\!\!\pmod{4}} t^{3n_1+3n_2} = -\frac{3t^6(1 + t^3)}{(1-t^6)(1-t^{12})},
\end{equation}
We must also subtract from \eqref{eq:vector1spin1} the contribution of all operators made by acting with any number of $\partial_{\pm\dot{+}}$ on $\partial_{+\dot{+}}\lambda^1_{-}-\partial_{-\dot{+}}\lambda^1_{+}=\partial_{\alpha\dot{+}}\lambda^{1\alpha}$. Those operators are identically zero by the Dirac equation. Since $\partial_{\alpha\dot{+}}\lambda^{1\alpha}$ is invariant under $\Gamma$, we can account for this correction by simply subtracting $-t^6$ from $I_\Phi$, according to the charges of $\partial_{\alpha\dot{+}}\lambda^{1\alpha}$ in \tabref{tab:SingleLetter}.
Adding~\eqref{eq:vector1spin0}, with $I_\Phi$ replaced by $I_\Phi+t^6$, and~\eqref{eq:vector1spin1}, we find that the vector multiplet contribution to the single letter index for holonomies 1-4 is 
\begin{align*}
    I^{vec}_1
    &=\frac{3(1-t^6+t^{12})(t^2-t^4+2t^6)- 3(1+t^3)t^6}{(1-t^6)(1-t^{12})}.
\end{align*}

Next, we compute $I^{vec}_5$, which is the vector multiplet single letter index for the $5^{th}$ holonomy of \tabref{tab:RepsDic}. This case is harder because the $5^{th}$ holonomy is nonabelian. 
The transformations of the adjoint field components $\Phi_p$ ($p=-1,0,1$) under the $SU(2)$ gauge transformations $g_5(r)$  and $g_5(s)$ are given in \eqref{eq:vectormultipletholonomyaction1}-\eqref{eq:vectormultipletholonomyaction2}, which can be written more concisely as
$$
g_5(r)\Phi_p g_5(r)^{-1}=(-1)^p\Phi_p\,,\qquad
g_5(s)\Phi_p g_5(s)^{-1}=-\Phi_{-p}\,.
$$
As before, we perform the computation for fields with spin projection $j_1=0$ and $j_1=\pm 1/2$ separately. A generic operator that corresponds to a spin $j_1=0$ field has the form:


\[
\sum c^{p n_1 n_2} \partial^{n_1}_{+ \dot{+}} \partial^{n_2}_{- \dot{+}} \Phi_p.  
\]
The $r$ and $s$ constraints are
\begin{align*}
   \sum e^{i\pi(n_1-n_2)/2} c^{p n_1 n_2} \partial^{n_1}_{+ \dot{+}} \partial^{n_2}_{- \dot{+}} \Phi_p  &= \sum (-1)^p c^{p n_1 n_2} \partial^{n_1}_{+ \dot{+}} \partial^{n_2}_{- \dot{+}}  \Phi_p,  \\
   \sum (-1)^{n_2} c^{p n_1 n_2} \partial^{n_2}_{+ \dot{+}} \partial^{n_1}_{- \dot{+}} \Phi_p   &= -\sum  c^{p n_1 n_2} \partial^{n_1}_{+ \dot{+}} \partial^{n_2}_{- \dot{+}} \Phi_{-p}. 
\end{align*}
They imply
\begin{equation}\label{eqn:ConstraintsHol5}
e^{i\pi(n_1-n_2)/2} c^{p n_1 n_2} = (-1)^p c^{p n_1 n_2}, \qquad
(-1)^{n_1} c^{-p, n_2 n_1} = -c^{p n_1 n_2}. 
\end{equation}
The second constraint expresses $c^{(-1)n_1 n_2}$ in terms of $c^{1n_2 n_1}$, so we only need to consider $p=0,1$. We now perform the counting for $p=0$ and $p=1$ separately, and then add up the two contributions.
\begin{itemize}
\item
 For $p=0$, the leftmost constraint of \eqref{eqn:ConstraintsHol5} requires
\[
n_1 - n_2 \equiv 0\pmod{4},
\]
and the second constraint expresses $c^{p n_1 n_2}$ for $n_1<n_2$ in terms of $c^{p n_2 n_1}$, and restricts $n_1$ to be odd when $n_1=n_2$. Thus, the number of independent parameters $c^{p n_1 n_2}$ is the number of pairs $(n_1, n_2)$ with $n_1>n_2$ and $n_1\equiv n_2\pmod{4}$ plus the number of odd $n_1$'s. 
The contribution of those to the sum is
\begin{equation}
F^0 = \frac{t^6}{(1-t^6)(1-t^{12})}\left\lbrack I_\Phi(t)-(-t^6)\right\rbrack.
\end{equation}
Here, we have already subtracted the term $(-t^6)$ corresponding to the contribution of the Dirac equation $\partial_{\alpha\dot{+}}\lambda^{1\alpha}=0$, in order to not over count operators constructed from derivatives of terms with $j_1=\pm\frac{1}{2}$ (which will be counted later on).
\item
For $p=1$, the leftmost constraint of \eqref{eqn:ConstraintsHol5} shows that
\[
n_1 - n_2 \equiv 2\pmod{4}.
\]
The contribution to the sum is therefore
\begin{equation}
F^+ = \frac{2t^6}{(1-t^6)(1-t^{12})}\left\lbrack I_\Phi(t)-(-t^6)\right\rbrack = 2F^0.
\end{equation}
\end{itemize}
Overall, the contribution to the single letter vector multiplet index for holonomy 5 from fields that have spin projection $j_1=0$ is
\begin{equation}
\label{eq:vector5spin0}
F^0 + F^+ = 
\frac{3t^6(t^2- t^4+2t^6)}{(1-t^6)(1-t^{12})}.
\end{equation}
Now let us consider the $j_1=\pm 1/2$ contribution. A generic operator of this kind takes the form
\[
\sum c^{p\mu n_1n_2} \partial^{n_1}_{+ \dot{+}} \partial^{n_2}_{- \dot{+}} \lambda_{\mu p},  
\]
where we recall that $\mu=\pm 1$ labels the $j_1=\mu/2$ and $p=-1,0,1$ labels the $\mathfrak{su}(2)$ Lie algebra component. The constraints are
\begin{align*}
\sum e^{i \pi(n_1-n_2 +\mu)/2} c^{p\mu n_1n_2} \partial^{n_1}_{+ \dot{+}} \partial^{n_2}_{- \dot{+}} \lambda_{\mu p}  &= \sum (-1)^p c^{p\mu n_1n_2}\partial^{n_1}_{+ \dot{+}} \partial^{n_2}_{- \dot{+}} \lambda_{\mu p},   \\
\sum (-1)^{n_2+(\mu-1)/2}c^{p\mu n_1n_2}\partial^{n_2}_{+ \dot{+}} \partial^{n_1}_{- \dot{+}} \lambda_{-\mu p} &= -\sum  c^{p\mu n_1n_2} \partial^{n_1}_{+ \dot{+}} \partial^{n_2}_{- \dot{+}} \lambda_{\mu, -p}.  
\end{align*}
They imply 
\begin{equation}
\label{eqn:ConstraintsHolSpinors5}
     e^{i\pi(n_1-n_2 +\mu)/2} c^{p\mu n_1n_2} = (-1)^p c^{p\mu n_1n_2}, \qquad
     (-1)^{n_1-(\mu+1)/2}c^{-p,-\mu, n_2n_1} = -c^{p \mu n_1n_2}.
\end{equation}
The second equation of \eqref{eqn:ConstraintsHolSpinors5} expresses $c^{(-1)\mu n_1 n_2}$ in terms of $c^{1(-\mu) n_2 n_1}$, and so we only need to count independent coefficients $c^{p\mu n_1 n_2}$ for $p=0,1$. We separate the calculation into four cases.
\begin{itemize}
\item
For $p=0, \mu=1$, the first constraint of \eqref{eqn:ConstraintsHolSpinors5} requires
\[
n_1 -n_2 \equiv 3\pmod{4}.
\]
The generating function for the number of such pairs $(n_1, n_2)$, each pair counted as $t^{3(n_1+n_2)}$, is
$$
\frac{t^3+t^9}{(1-t^6)(1-t^{12})},
$$
and the contribution to the index is
\begin{equation}
F^{01} = -\frac{t^3(t^3+t^9)}{(1-t^6)(1-t^{12})}.
\end{equation}
where the prefactor $(-t^3)$ corresponds to the quantum numbers of $\lambda^1_\mu$, as listed in \tabref{tab:SingleLetter}.

\item 
For $p=0, \mu=-1$, since the second constraint of \eqref{eqn:ConstraintsHolSpinors5} determines $c^{0(-1)n_1 n_2}$ in terms of $c^{0 1 n_2 n_1}$, there are no additional contributions to the index that have not been accounted for by the previous case, $p=0,\mu=1$.

\item
For $p=1, \mu=1$, the first constraint of \eqref{eqn:ConstraintsHolSpinors5} implies
\[
n_1 - n_2 \equiv 1\pmod{4}.
\]
The sum is therefore
\begin{equation}
F^{11} = -\frac{t^3(t^3+t^9)}{(1-t^6)(1-t^{12})} = F^{01}.
\end{equation}

\item
For $p=1, \mu=-1$, the first constraint of \eqref{eqn:ConstraintsHolSpinors5} implies 
\[
n_1 - n_2 \equiv 3\pmod{4}.
\]
The sum is therefore
\begin{equation}
F^{1,-1} = -\frac{t^3(t^3+t^9)}{(1-t^6)(1-t^{12})} = F^{01}.
\end{equation}
\end{itemize}

Adding all four contributions together, we find that the $j_1=\pm 1/2$ contribution to the single letter vector multiplet index is
\begin{equation}
\label{eq:vector5spin1}
F^{01}+F^{11}+F^{1,-1} = -\frac{3t^3(t^3+t^9)}{(1-t^6)(1-t^{12})}.
\end{equation}
Adding together the $j_1=0$ contribution from \eqref{eq:vector5spin0} and the $j_1=\pm 1/2$ contribution from \eqref{eq:vector5spin1}, we get the vector multiplet contribution to the index for holonomy 5:
\begin{align*}
I^{vec}_5 &=
-3\frac{t^6(1+t^4)(1-t^2)}{(1-t^6)(1-t^{12}) }.
\end{align*}
In summary, we have the following results for the vector multiplet contribution to the single letter index:
\begin{align}
    \label{eq:vector1index}
    I^{vec}_a &=
    \frac{3t^2(1-t^6+t^{12})(1-t^2+2t^4)- 3(t^3 + t^9)t^3}{(1-t^6)(1-t^{12})},
    \qquad (a=1,\dots,4),\\
    \label{eq:vector5index}
    I^{vec}_5 &=
-3\frac{t^6(1+t^4)(1-t^2)}{(1-t^6)(1-t^{12}) }.
\end{align}

Now, let us move on to compute the trifundamental contribution to the single letter index.


\subsection{The contribution $I^{\text{(trifund)}}$ of the half-hypermultiplet}
We go over the explicit calculation of the single letter index $I^{\text{(trifund)}}_{abc}$ associated to the trifundamental field, where $a,b,c \in \{1,2,3,4\}$. According to equation~\eqref{eq:hypermultipletholonomyaction} and \tabref{tab:RepsDic}, there are only 4 independent combinations of $(abc)$ that give unique constraints, which we can take to be $(111)$, $(112)$, $(113)$, $(114)$, for example. The holonomy triplet $(122)$, for instance, is equivalent to $(111)$ in terms of the implied constraints on the trifundamental fields, since the factors $g_a(r)g_b(r)g_c(r)$ and $g_a(s)g_b(s)g_c(s)$ on the left hand side of \eqref{eq:hypermultipletholonomyaction} are the same for both $(a,b,c)=(1,1,1)$ and $(a,b,c)=(1,2,2)$. Therefore, we only have 4 independent indices to compute. We also note that all component fields in the half-hypermultiplet have $j_1=0$, so we need not worry about the $j_1=\pm 1/2$ contributions.  

In the case of $I^{\text{(trifund)}}_{111}$, a generic operator that arises from the state-operator correspondence now has the form:
\[
\sum c^{n_1n_2} \partial^{n_1}_{+ \dot{+}} \partial^{n_2}_{- \dot{+}} \Phi,  
\]
where $\Phi$ denotes a trifundamental field. For clarity, we suppress the SU$(2)$ trifundamental indices on the field $\Phi$, which otherwise would appear as in equation~\eqref{eq:hypermultipletholonomyaction}. The $r$ and the $s$ constraints give
\begin{equation}
\label{eqn:ccrstri}
    c^{n_1n_2}=e^{i\pi(n_1-n_2)/2}c^{n_1n_2},
    \qquad
    c^{n_1n_2}=(-1)^{n_1}c^{n_2n_1}.
\end{equation}
The first constraint implies that 
\[
n_1 -n_2\equiv 0\pmod{4},
\]
and the second constraint of \eqref{eqn:ccrstri} shows that the coefficients $c^{n_1n_2}$ for $n_1<n_2$ are determined in terms of the coefficients $c^{n_1 n_2}$ for $n_1>n_2$, and for $n_1=n_2$ only the coefficients with even $n_1$ can be nonzero. Overall, we find that the number of independent coefficients $c^{n_1 n_2}$ with a given $n_1+n_2$ is the coefficient of $t^{3(n_1+n_2)}$ in
$$
\frac{1-t^6+t^{12}}{(1-t^6)(1-t^{12})}\,.
$$
Altogether, the trifundamental index is
\begin{equation}
I^{\text{(trifund)}}_{111} =8\frac{(1-t^6+t^{12})(t^2-t^4)}{(1-t^6)(1-t^{12})}.
\end{equation}
where the prefactor 8 comes from $2^3$ choices of the SU$(2)$ trifundamental indices, and the factor $(t^2-t^4)$ comes from the contribution of the fields $q^1$ and $\overline{\psi}_{\dot{+}}$ as listed in \tabref{tab:SingleLetter}, with $v=1$.

In the case of $I^{\text{(trifund)}}_{112}$, the constraints are
\begin{equation}
    e^{i\pi(n_1-n_2)/2}c^{n_1n_2} =c^{n_1n_2},
    \qquad
    (-1)^{n_1}c^{n_2n_1} =-c^{n_1n_2},
\end{equation}
and the number of pairs $(n_1, n_2)$ with given $n_1+n_2$ is given by the coefficient of $t^{3(n_1+n_2)}$ in the expansion of
$$
\frac{t^6}{(1-t^6)(1-t^{12})}\,,
$$
and so
\begin{equation}
I^{\text{(trifund)}}_{112} 
=8\frac{t^6(t^2-t^4)}{(1-t^6)(1-t^{12})}.
\end{equation}

In the case of $I^{\text{(trifund)}}_{113}$, the constraints are
\begin{equation}
    e^{i\pi(n_1-n_2)/2}c^{n_1n_2} =-c^{n_1n_2}, \qquad
    (-1)^{n_1}c^{n_2n_1} =c^{n_1n_2}.
\end{equation}
The first constraint implies 
\[
n_1-n_2 \equiv 2\pmod{4},
\]
and the second one tells us to count all such $n_1,n_2$ pairs with $n_1>n_2$. Their number, for a given $n_1+n_2$, is given by the coefficient of $t^{3(n_1+n_2)}$ in the expansion of
$$
\frac{t^6}{(1-t^6)(1-t^{12})},
$$
as we found in the computation of $I^{\text{(trifund)}}_{112}$ above.
This gives
\begin{equation}
I^{trifund}_{113} 
=8\frac{t^6(t^2-t^4)}{(1-t^6)(1-t^{12})}.
\end{equation}

In the case of $I^{\text{(trifund)}}_{114}$, the constraints are
\begin{align*}
    e^{i\pi(n_1-n_2)/2}c^{n_1n_2} =-c^{n_1n_2},
    \qquad
    (-1)^{n_1}c^{n_2n_1} =-c^{n_1n_2}.
\end{align*}
Comparing the (114) constraints with the (113) constraints, we see that they have the same dimension of the space of solutions, so
\begin{equation}
I^{\text{(trifund)}}_{114} = I^{\text{(trifund)}}_{113}=8\frac{t^6(t^2-t^4)}{(1-t^6)(1-t^{12})}.
\end{equation}
This completes the computation for the case when all three holonomies are abelian. 

Now, we study the index when one, two, or three of the holonomies in the triplet are the 5${}^{th}$ holonomy. First, consider the case where there is only one non-abelian holonomy. Similarly to the argument we had previously, there are only 4 independent cases to consider: (115), (125), (135), (145). 

In the case of $I^{\text{(trifund)}}_{115}$, the SU$(2)$ index of the third component of the trifundamental field matters, because the nonabelian holonomy does not act diagonally anymore. This consideration leads to the most general form of the operator:
\[
\sum c^{\mu n_1n_2} \partial^{n_1}_{+ \dot{+}} \partial^{n_2}_{- \dot{+}} \Phi_\mu,  
\]
where $\mu= \pm 1$ is the gauge index of the trifundamental field with respect to the third SU$(2)$ factor. For clarity, we suppress the gauge indices of the first two SU$(2)$ factors. The constraints are now
\begin{align*}
   \sum e^{i\pi(n_1-n_2)/2}c^{\mu n_1n_2} \partial^{n_1}_{+ \dot{+}} \partial^{n_2}_{- \dot{+}} \Phi_\mu   &=
   \sum e^{\mu\pi i/2}c^{\mu n_1n_2} \partial^{n_1}_{+ \dot{+}} \partial^{n_2}_{- \dot{+}} \Phi_\mu,  \\
   \sum (-1)^{n_2}c^{\mu n_1n_2}\partial^{n_2}_{+ \dot{+}} \partial^{n_1}_{- \dot{+}} \Phi_\mu  &=
   \sum (-1)^{(\mu-1)/2}c^{\mu n_1n_2} \partial^{n_1}_{+ \dot{+}} \partial^{n_2}_{- \dot{+}} \Phi_{-\mu}, 
\end{align*}
which lead to the equations
\begin{equation}\label{eqn:115cctri}
    e^{i\pi(n_1-n_2)/2}c^{\mu n_1n_2} = e^{\mu\pi i/2}c^{\mu n_1n_2}, \qquad
    (-1)^{n_1}c^{-\mu, n_2n_1} = (-1)^{(\mu-1)/2}c^{\mu n_1n_2}.
\end{equation}
For $\mu=1$, the first constraint of \eqref{eqn:115cctri} implies 
\begin{align*}
    n_1-n_2 \equiv 1\pmod{4},
\end{align*}
and the second constraint instructs us to count all such $n_1,n_2$ pairs with $\mu=1$, since the $\mu=-1$ coefficients are determined by the $\mu=1$ ones. For given $n_1+n_2$, this number is the coefficient of $t^{3(n_1+n_2)}$ in the expansion of
$$
\frac{t^3+ t^9}{(1-t^6)(1-t^{12})},
$$
and the contribution to the index is
\begin{equation}
I^{\text{(trifund)}}_{115} 
=  4\frac{(t^3+ t^9)(t^2-t^4)}{(1-t^6)(1-t^{12})},
\end{equation}
where the $4$ comes from the $2^2$ choices of the first two SU$(2)$ trifundamental indices. The factor $t^2-t^4$ comes from the contribution of the half-hypermultiplet base fields $q^1$ and $\bar{\psi}_{\dot{+}}$.

In the case of $I^{\text{(trifund)}}_{125}$, the constraints are 
\begin{align*}
   \sum e^{i\pi(n_1-n_2)/2}c^{\mu n_1n_2} \partial^{n_1}_{+ \dot{+}} \partial^{n_2}_{- \dot{+}} \Phi_\mu   &=
   \sum e^{\mu\pi i/2}c^{\mu n_1n_2} \partial^{n_1}_{+ \dot{+}} \partial^{n_2}_{- \dot{+}} \Phi_\mu,   \\
   \sum (-1)^{n_2}c^{\mu n_1n_2} \partial^{n_2}_{+ \dot{+}} \partial^{n_1}_{- \dot{+}} \Phi_\mu   &=
   -\sum (-1)^{(\mu-1)/2}c^{\mu n_1n_2}\partial^{n_1}_{+ \dot{+}} \partial^{n_2}_{- \dot{+}}\Phi_{-\mu}. 
\end{align*}
The only difference from the previous $115$ case is the minus sign for the $s$ condition, but this sign does not change the count, and
\begin{equation}
I^{\text{(trifund)}}_{125} = I^{\text{(trifund)}}_{115} =  4\frac{(t^3+ t^9)(t^2-t^4)}{(1-t^6)(1-t^{12})}.
\end{equation}
In the case of $I^{\text{(trifund)}}_{135}$, the constraints are
\begin{align*}
   \sum e^{i\pi(n_1-n_2)/2}c^{\mu n_1n_2}\partial^{n_1}_{+ \dot{+}} \partial^{n_2}_{- \dot{+}} \Phi_\mu  &=
   -\sum e^{\mu\pi i/2}c^{\mu n_1n_2} \partial^{n_1}_{+ \dot{+}} \partial^{n_2}_{- \dot{+}} \Phi_\mu,   \\
   \sum (-1)^{n_2}c^{\mu n_1n_2}\partial^{n_2}_{+ \dot{+}} \partial^{n_1}_{- \dot{+}}\Phi_\mu  &=
   \sum (-1)^{(\mu-1)/2}c^{\mu n_1n_2}\partial^{n_1}_{+ \dot{+}} \partial^{n_2}_{- \dot{+}} \Phi_{-\mu}.  
\end{align*}
Now, the $\mu=1$ case leads to
\[
n_1 - n_2 \equiv 3\pmod{4},
\]
and it is easy to see that 
\begin{equation}
I^{\text{(trifund)}}_{135} = I^{\text{(trifund)}}_{115} =  4\frac{(t^3+ t^9)(t^2-t^4)}{(1-t^6)(1-t^{12})}.
\end{equation}
By the same logic, we have
\begin{equation}
I^{\text{(trifund)}}_{145} = I^{\text{(trifund)}}_{115} =  4\frac{(t^3+ t^9)(t^2-t^4)}{(1-t^6)(1-t^{12})}.
\end{equation}
This completes the computation for the cases where only one of the three holonomies is nonabelian. We now consider the four cases where two of the holonomies are nonabelian. Those are: (155), (255), (355), and (455).

In the case of $I^{trifund}_{155}$, the second and the third SU$(2)$ trifundamental gauge indices matter, since the holonomies will act non-diagonally on them separately. Suppressing only the gauge indices of the first SU$(2)$ factor, we write the components of the trifundamental field as $\Phi_{\mu\nu}$ ($\mu,\nu=\pm 1$), and the general operator combination is written as
\[
\sum c^{\mu\nu n_1n_2} \partial^{n_1}_{+ \dot{+}} \partial^{n_2}_{- \dot{+}} \Phi_{\mu\nu}.  
\]
The constraints are
\begin{align*}
    \sum e^{i \pi (n_1-n_2)/2}c^{\mu\nu n_1n_2} \partial^{n_1}_{+ \dot{+}} \partial^{n_2}_{- \dot{+}} \Phi_{\mu\nu}  &= \sum e^{i\pi (\mu+\nu)/2}c^{\mu\nu n_1n_2}  \partial^{n_1}_{+ \dot{+}} \partial^{n_2}_{- \dot{+}} \Phi_{\mu\nu}, \\
    \sum (-1)^{n_2}c^{\mu\nu n_1n_2}\partial^{n_2}_{+ \dot{+}} \partial^{n_1}_{- \dot{+}}  \Phi_{\mu\nu}  &= \sum (-1)^{(\mu+\nu)/2-1}c^{\mu\nu n_1n_2}\partial^{n_1}_{+ \dot{+}} \partial^{n_2}_{- \dot{+}}  \Phi_{-\mu,-\nu}, 
\end{align*}
from which we obtain
\begin{equation}
\label{eqn:cmunun1n2}
    e^{i\pi(n_1-n_2)/2}c^{\mu\nu n_1n_2}=e^{i\pi (\mu+\nu)/2}c^{\mu\nu n_1n_2}, \quad
    (-1)^{n_1}c^{-\mu,-\nu, n_2n_1} = (-1)^{(\mu+\nu)/2-1}c^{\mu\nu n_1n_2}.
\end{equation}
We proceed by counting the number of linearly independent $c^{\mu\nu n_1 n_2}$'s for each $(\mu,\nu)$ separately.

\begin{itemize}
\item
For $\mu=\nu=1$, the first equation of \eqref{eqn:cmunun1n2} implies
\begin{equation}
\label{eqn:n1n2equiv2}
n_1-n_2 \equiv 2\pmod{4},    
\end{equation}
and the second equation says that the coefficients with $\mu=\nu=-1$ are determined by those with $\mu=\nu=1$. The number of pairs satisfying \eqref{eqn:n1n2equiv2} for a given $n_1+n_2$ is the coefficient of $t^{3(n_1+n_2)}$ in the expansion of
$$
W_{11}=\frac{2t^6}{(1-t^6)(1-t^{12})}.
$$
\item
For $\mu=1, \nu=-1$, the first constraint of \eqref{eqn:cmunun1n2} implies
\begin{equation}
\label{eqn:n1n2equiv0}
n_1-n_2 \equiv 0\pmod{4},    
\end{equation}

and the second constraint determines the coefficients with $\mu=-1$ and $\nu=1$ in terms of those with $\mu=1$ and $\nu=-1$.
The generating function is
\[
W_{1,-1} = \frac{1+t^{12}}{(1-t^6)(1-t^{12})}. 
\]
\end{itemize}
Adding these together, we have
\begin{equation}
I^{\text{(trifund)}}_{155} =2(W_{11}+W_{1,-1})(t^2-t^4) = \frac{2(1+2t^6+t^{12})(t^2-t^4)}{(1-t^6)(1-t^{12})},
\end{equation}
where the prefactor 2 comes from 2 choices of the first SU$(2)$ trifundamental gauge index, and $t^2-t^4$ is the contribution of the fields $q^1$ and $\bar{\psi}_{\dot{+}}$.

In the case of $I^{\text{(trifund)}}_{255}$, the constraints are
\begin{align*}
    \sum e^{i\pi(n_1-n_2)/2}c^{\mu\nu n_1n_2} \partial^{n_1}_{+ \dot{+}} \partial^{n_2}_{- \dot{+}} \Phi_{\mu\nu}  &= \sum e^{i\pi (\mu+\nu)/2}c^{\mu\nu n_1n_2} \partial^{n_1}_{+ \dot{+}} \partial^{n_2}_{- \dot{+}} \Phi_{\mu\nu},   \\
    \sum (-1)^{n_2}c^{\mu\nu n_1n_2}\partial^{n_2}_{+ \dot{+}} \partial^{n_1}_{- \dot{+}}  \Phi_{\mu\nu}  &= -\sum (-1)^{(\mu+\nu)/2-1}c^{\mu\nu n_1n_2} \partial^{n_1}_{+ \dot{+}} \partial^{n_2}_{- \dot{+}} \Phi_{-\mu,-\nu} . 
\end{align*}
The only difference from the previous (155) case is the minus sign in the second equation, but this does not affect the number of independent solutions. Thus,
\begin{equation}
I^{\text{(trifund)}}_{255} =I^{\text{(trifund)}}_{155} = \frac{2(1+2t^6+t^{12})(t^2-t^4)}{(1-t^6)(1-t^{12})}.
\end{equation}
In the case of $I^{\text{(trifund)}}_{355}$, the constraints are
\begin{align*}
    \sum e^{i\pi(n_1-n_2)/2}c^{\mu\nu n_1n_2} \partial^{n_1}_{+ \dot{+}} \partial^{n_2}_{- \dot{+}} \Phi_{\mu\nu}  &= -\sum e^{i\pi (\mu+\nu)/2}c^{\mu\nu n_1n_2} \partial^{n_1}_{+ \dot{+}} \partial^{n_2}_{- \dot{+}} \Phi_{\mu\nu},   \\
    \sum (-1)^{n_2}c^{\mu\nu n_1n_2}\partial^{n_2}_{+ \dot{+}} \partial^{n_1}_{- \dot{+}}  \Phi_{\mu\nu}  &= \sum (-1)^{(\mu+\nu)/2-1}c^{\mu\nu n_1n_2} \partial^{n_1}_{+ \dot{+}} \partial^{n_2}_{- \dot{+}} \Phi_{-\mu,-\nu},   
\end{align*}
from which we obtain
\begin{equation}
    e^{i \pi(n_1-n_2)/2}c^{\mu\nu n_1n_2}=-e^{i \pi (\mu+\nu)/2}c^{\mu\nu n_1n_2}, \qquad
    (-1)^{n_1}c^{-\mu,-\nu, n_2n_1} = (-1)^{(\mu+\nu)/2-1}c^{\mu\nu n_1n_2}.
\end{equation}
\begin{itemize}
\item
If $\mu=\nu=1$, the first constraint implies
\[
n_1-n_2 \equiv 0\pmod{4}.
\]
\item
If $\mu=1, \nu=-1$, the first constraint implies
\[
n_1-n_2 \equiv 2\pmod{4}.
\]
\end{itemize}
Noting that these are for the same constraints as the (155) case [equations \eqref{eqn:n1n2equiv2} and \eqref{eqn:n1n2equiv0}] except that \eqref{eqn:n1n2equiv2} goes with $\nu=-1$ this time and \eqref{eqn:n1n2equiv0} goes with $\nu=1$, we immediately get
\begin{equation}
I^{\text{(trifund)}}_{355} =I^{\text{(trifund)}}_{155} = \frac{2(1+2t^6+t^{12})(t^2-t^4)}{(1-t^6)(1-t^{12})}.
\end{equation}
By the same logic,
\begin{equation}
I^{\text{(trifund)}}_{455} =I^{\text{(trifund)}}_{155} = \frac{2(1+2t^6+t^{12})(t^2-t^4)}{(1-t^6)(1-t^{12})}.
\end{equation}
In summary, for the nonabelian holonomies that appear twice in the triplet, we find
\begin{equation}
I^{\text{(trifund)}}_{155}=I^{\text{(trifund)}}_{255}=I^{\text{(trifund)}}_{355}=I^{\text{(trifund)}}_{455} = \frac{2(1+2t^6+t^{12})(t^2-t^4)}{(1-t^6)(1-t^{12})}.
\end{equation}

Finally, we need to consider the last case, holonomy (555). A general operator is 
\[
\sum c^{\mu\nu\rho n_1n_2} \partial^{n_1}_{+ \dot{+}} \partial^{n_2}_{- \dot{+}} \Phi_{\mu\nu\rho},  
\]
where now all three trifundamental SU$(2)$ indices ($\mu,\nu,\rho =\pm 1$) are important. The constraints are
\begin{align*}
 \sum e^{i\pi(n_1-n_2)/2} c^{\mu\nu\rho n_1n_2}\partial^{n_1}_{+ \dot{+}} \partial^{n_2}_{- \dot{+}} \Phi_{\mu\nu\rho} &= \sum e^{i\pi (\mu+\nu+\rho)/2} c^{\mu\nu\rho n_1n_2} \partial^{n_1}_{+ \dot{+}} \partial^{n_2}_{- \dot{+}} \Phi_{\mu\nu\rho},    \\
 \sum (-1)^{n_2} c^{\mu\nu\rho n_1n_2} \partial^{n_2}_{+ \dot{+}} \partial^{n_1}_{- \dot{+}} \Phi_{\mu\nu\rho}  &= \sum (-1)^{(\mu+\nu+\rho-3)/2} c^{\mu\nu\rho n_1n_2} \partial^{n_1}_{+ \dot{+}} \partial^{n_2}_{- \dot{+}} \Phi_{-\mu,-\nu,-\rho},   
\end{align*}
from which we derive
\begin{align*}
    e^{i\pi(n_1-n_2)/2} c^{\mu\nu\rho n_1n_2} &=e^{i\pi(\mu+\nu+\rho)/2} c^{\mu\nu\rho n_1n_2}, \\
    (-1)^{n_1} c^{-\mu,-\nu,-\rho n_2n_1} &=(-1)^{(\mu+\nu+\rho-3)/2} c^{\mu\nu\rho n_1n_2}.
\end{align*}

\begin{itemize}

\item
For $\mu=1, \nu=1, \rho=1$, we have
$n_1-n_2 \equiv 3\pmod{4}.$
\item 
For $\mu=1, \nu=1, \rho=-1$, we have
$n_1-n_2 \equiv 1\pmod{4}$.
\item
For $\mu=1, \nu=-1, \rho=1$, we have
$n_1-n_2 \equiv 1\pmod{4}.$
\item 
For $\mu=1, \nu=-1, \rho=-1$, we have
$n_1-n_2 \equiv 3\pmod{4}.$

\end{itemize}
For each of the above four cases, the number of pairs $n_1, n_2$ that satisfy the mod 4 constraint, with given $n_1+n_2$, is the coefficient of $t^{3(n_1+n_2)}$ in the expansion of
\[
\frac{t^3+t^9}{(1-t^6)(1-t^{12})}.
\]
Adding these up, and including the $t^2-t^4$ contribution from the half-hypermultiplet base fields, we have
\begin{equation}
I^{\text{(trifund)}}_{555} = 4\frac{(t^3+t^9)(t^2-t^4)}{(1-t^6)(1-t^{12})}.
\end{equation}
Now, we have all the components we need to compute the supersymmetric Casimir energy.

\subsection{Supersymmetric Casimir Energy computation}
When we sum over all the indices to compute the Casimir energy, we need to multiply the vector multiplet contribution by a factor of $1/2$, since each vector multiplet is shared between two Trinions.
We also set $t=\exp\beta$ and expand near $\beta=0$. For example, for the triplet (111) we have
\begin{align*}
    I_{111} &= I^{\text{(trifund)}}_{111} + \frac{3}{2} I^{\text{(vec)}}_{1}
=-\frac{1}{18\beta} +\frac{9}{2} - \frac{35\beta}{54} + O(\beta^2),
\end{align*}
which leads, using \eqref{eqn:CasimirDef}, to the supersymmetric Casimir energy contribution from the (111) vertex,
\begin{equation}
E_{111} = \frac{35}{108}.
\end{equation}
It follows that 
\begin{equation}
E_{111}=E_{122} =E_{133} =E_{144} = E_{234} =\frac{35}{108},
\end{equation}
since these holonomies have the same index contributions as we computed in the previous subsection.

For the other holonomies that do not involve holonomy 5:
\begin{align}
    I_{112} &= I^{\text{(trifund)}}_{112} + \frac{1}{2} \left(2I^{\text{(vec)}}_{1} 
    +I^{\text{(vec)}}_{1}\right)
    =-\frac{1}{18\beta} +\frac{9}{2} - \frac{251\beta}{54} + O(\beta^2), \\
    I_{114}&=I_{113} = I^{\text{(trifund)}}_{113} +\frac{1}{2} \left(2I^{\text{(vec)}}_{1} 
    +I^{\text{(vec)}}_{3}\right)
    =-\frac{1}{18\beta} +\frac{9}{2} - \frac{251\beta}{54} + O(\beta^2).
\end{align}
These show that
\begin{align}
E_{112}&= E_{113} = E_{114} = E_{222}= E_{332} =E_{442} = E_{134}= \frac{251}{108}, \\
E_{224}&=E_{334} = E_{444}=E_{223}=E_{333}=E_{443}=E_{123}=E_{124}= \frac{251}{108}.
\end{align}
For holonomies that involve a single nonabelian holonomy, we have
\begin{align}
    I_{115} &= I^{\text{(trifund)}}_{115} +  I^{\text{(vec)}}_{1}+\frac{1}{2} I^{\text{(vec)}}_{5} 
    =-\frac{1}{18\beta} +3 - \frac{143\beta}{54} + O(\beta^2),
\end{align}
which shows that
\begin{equation}
E_{115}=E_{225} =E_{335} = E_{445} = \frac{143}{108}.
\end{equation}

Similarly, one can show that $I_{125}=I_{135}=I_{145}=I_{115}$, so
\begin{equation}
E_{ab5} = E_{115} = \frac{143}{108},
\end{equation}
where $a,b\in\{1,2,3,4\}$. Holonomy triplets that involve only one nonabelian holonomy have the same Casimir energy.

Now, consider holonomy triplets that contain two nonabelian holonomies. 
\begin{align}
    I_{155} &= I^{\text{(trifund)}}_{155} +  \frac{1}{2}I^{\text{(vec)}}_{1}+I^{\text{(vec)}}_{5} 
    =-\frac{1}{18\beta} +\frac{3}{2} - \frac{35\beta}{54} + O(\beta^2).
\end{align}
What is surprising is that
\begin{equation}
E_{155} = \frac{35}{108} = E_{111}.
\end{equation}
From the previous section, we see that
\begin{equation}
E_{255} = E_{355} =E_{455} =E_{155} =\frac{35}{108}.
\end{equation}

The results we have so far, ranked from lowest energy to highest, is 
\begin{itemize}
\item $E_0=35/108$: (111), (122), (133), (144), (234), (155), (255), (355), (455). These lead to 28 states, and they correspond to the states in the dual $SO(8)$ Chern-Simons theory at level 2.

\item $E_1=143/108 = E_0+1$: $(ab5)$, $a,b\in\{1,2,3,4\}$. These lead to 48 states. 

\item $E_2= 251/108 = E_0+2$: (112), (113), (114), (332), (442), (222), (134), (224), (334), (223), (443), (333), (444), (123), (124). These lead to 48 states.
\end{itemize}
There is one more state, (555), that is left to consider. 
\begin{align}
    I_{555} &= I^{\text{(trifund)}}_{555} +  \frac{3}{2}I^{\text{(vec)}}_{5}
    =-\frac{1}{18\beta} +\frac{19}{54}\beta + O(\beta^2).
\end{align}
This case is special as it does not have a constant term in the single letter index, and the supersymmetric Casimir energy is negative:
\begin{equation}
E_3=E_{555} = -\frac{19}{108} = E_0 - \frac{1}{2}.
\end{equation}


\bibliographystyle{JHEP}
\bibliography{references.bib}

\end{document}